\documentclass[12pt,aps,floats,showpacs,showkeys,preprintnumbers,nofootinbib]{revtex4}

\usepackage{amsmath,amssymb,bbm}
\usepackage{array}
\usepackage{graphicx}
\usepackage[pdftex,colorlinks=false,urlcolor=blue,linkcolor=blue]{hyperref}
\usepackage{epstopdf}
\usepackage{slashed,xcolor}
\newcommand{\PreserveBackslash}[1]{\let\temp=\\#1\let\\=\temp}
\newcolumntype{C}[1]{>{\PreserveBackslash\centering}p{#1}}
\newcolumntype{R}[1]{>{\PreserveBackslash\raggedleft}p{#1}}
\newcolumntype{L}[1]{>{\PreserveBackslash\raggedright}p{#1}}
\newcommand{\sgn}{\mathop{\mathrm{sgn}}}

\def\tanb{\tan\beta}

\def\sina{\sin\alpha}

\def\sbma{\sin(\beta-\alpha)}
\def\sbpa{\sin(\beta+\alpha)}
\def\cbma{\cos(\beta-\alpha)}
\def\gam{\gamma}
\def\lam{\lambda}

\def\hl{h}
\def\hh{H}
\def\ha{A}

\def\hpm{H^\pm}
\def\mhl{m_{\hl}}
\def\mhh{m_{\hh}}
\def\mha{m_\ha}
\def\mhpm{m_{\hpm}}

\def\mz{m_Z}

\def\mot{m_{12}}
\def\mhat{\hat m_{12}}

\def\CD{C_D}

\def\CV{C_V}
\def\CG{C_g}
\def\CP{C_\gamma}

\def\cd{\CD}
\def\cv{\CV}
\def\cg{\CG}
\def\cp{\CP}

\def\gev{~{\rm GeV}}

\def\fbi{~{\rm fb}^{-1}}
\def\pb{~{\rm pb}}

\def\br{{\rm BR}}
\def\anti{\overline}

\def\eg{{\it e.g.}}
\def\ie{{\it i.e.}}

\def\Eq#1{Eq.~(\ref{#1})}

\def\Fig#1{Fig.~\ref{#1}}
\def\beq{\begin{equation}}
\def\eeq{\end{equation}}
\def\bea{\begin{eqnarray}}
\def\eea{\end{eqnarray}}
\def\bit{\begin{itemize}}
\def\eit{\end{itemize}}
\def\ben{\begin{enumerate}}
\def\een{\end{enumerate}}
\def\rgghvv{\mu_{gg}^h(VV)}
\def\rgghgamgam{\mu_{gg}^h(\gam\gam)}

\def\rggHgamgam{\mu_{gg}^H(\gam\gam)}
\def\rvbfHgamgam{\mu_{\rm VBF}^H(\gam\gam)}

\def\rgghzz{\mu_{gg}^h(VV)}
\def\rggHzz{\mu_{gg}^H(VV)}

\def\lsim{\mathrel{\raise.3ex\hbox{$<$\kern-.75em\lower1ex\hbox{$\sim$}}}}
\def\gsim{\mathrel{\raise.3ex\hbox{$>$\kern-.75em\lower1ex\hbox{$\sim$}}}}

\def\ifmath#1{\relax\ifmmode #1\else $#1$\fi}
\def\ls#1{\ifmath{_{\lower1.5pt\hbox{$\scriptstyle #1$}}}}
\def\lss#1{\ifmath{^{\,\lower2.5pt\hbox{$\scriptstyle #1$}}}}
\def\lamT{\lambda_T}
\def\lamU{\lambda_U}

\def\lamA{\lambda_A}

\def\quarter{\ifmath{{\textstyle{1 \over 4}}}}
\def\half{\ifmath{{\textstyle{1 \over 2}}}}

\def\sb  {s_{\beta}}
\def\cb  {c_{\beta}}
\def\stwob  {s_{2\beta}}

\def\sthreeb{s_{3\beta}}

\def\cthreeb{c_{3\beta}}

\def\tanb{\tan\beta}

\def\sina{\sin\alpha}

\def\lamtil{\lam\ls{345}}
\def\lamhat{\widehat\lam}
\def\nicefrac#1#2{\hbox{${#1\over #2}$}}

\def\cbpa{\cos(\beta+\alpha)}

\def\sbmaii{s^2_{\beta-\alpha}}
\def\cbmaii{c^2_{\beta-\alpha}}
\def\typei{Type~I}
\def\typeii{Type~II}
\def\ghaa{g\ls{\hl\ha\ha} }
\def\brhaa{\br(h\to AA)}

\def\gyxx{g\ls{YXX}}

\begin{document}
\title{Light Higgs bosons in Two-Higgs-Doublet Models}

\author{Jeremy~Bernon$^{1}$}
\email[]{jeremy.bernon@lpsc.in2p3.fr}
\author{John F.~Gunion$^{2}$}
\email[]{jfgunion@ucdavis.edu}
\author{Yun~Jiang$^{2}$}
\email[]{yunjiang@ucdavis.edu}
\author{Sabine~Kraml$^{1}$}
\email[]{sabine.kraml@lpsc.in2p3.fr}
\affiliation{(1) \,Laboratoire de Physique Subatomique et de Cosmologie, Universit\'e Grenoble-Alpes,
CNRS/IN2P3, 53 Avenue des Martyrs, F-38026 Grenoble, France}
\affiliation{(2) \,Department of Physics, University of California, Davis, CA 95616, USA}

\begin{abstract}
We explore the possibilities in two-Higgs-doublet models (2HDMs) of Type~I and Type~II 
for Higgs states with mass below about 60~GeV, \ie\ less than half of the $\sim 125\gev$ mass 
of the observed SM-like Higgs boson. We identify the latter as either the lighter or the heavier CP-even 
state, $h$ or $H$, and employ scans of the 2HDM parameter space taking into account all relevant 
theoretical and experimental constraints, including the most up-to-date Higgs signal strength measurements. 
We find that, in both Type~I and Type~II models, such light Higgs states are phenomenologically viable and 
can lead to interesting signatures. Part of the relevant parameter space may be testable with the existing 8 TeV LHC data, 
\eg\ by looking for direct production of the light state via $gg$-fusion or $b\anti b$-associated-production using its $\tau^+\tau^-$ and $\mu^+\mu^-$ decays at low invariant mass.  
\end{abstract}

\pacs{12.60.Fr, 14.80.Ec, 14.80.Fd}
\keywords{Higgs physics, 2-Higgs-Doublet Model, LHC}
\preprint{LPSC14305\cr UCD-2014-004}

\maketitle

\clearpage
\section{Introduction}

Now that a new particle has been discovered at the LHC with properties close to those of the SM Higgs boson, 
it is important to assess all possibilities for other Higgs-like states that may have escaped detection at Run~1 of the LHC. 
Two-Higgs-doublet models (2HDMs --- we consider \typei\ and \typeii\ models) are an especially simple 
and appealing framework for such considerations.  
They contain five Higgs bosons (the CP-even $h$ and $H$, the CP-odd $A$, and the charged states $\hpm$) 
where the $h$ or $H$ can have SM-like couplings and may therefore be identified with the observed $125\gev$ state --- denoted $h125$ and $H125$, respectively. 
One often considered limit of the 2HDM is the decoupling limit~\cite{Gunion:2002zf} in which $\mha,\mhh,\mhpm$ 
are all large, in which case the $h$ is very SM-like.  

A SM-like $h$ or $H$ can however also be obtained in the 
alignment limit without the masses of the other Higgs being large.  
Here, we address the seemingly extreme case in which the $h$ ($H$) is the SM-like $125\gev$ state 
and the $\ha$ ($\ha$ and/or $\hl$) are lighter than $125\gev$, in particular light enough that 
the SM-like state can decay into them.  Such decays generically have a large branching ratio (early references are \cite{Gunion:1984yn}, \cite{Li:1985hy} and \cite{Gunion:1989we})
and would 
conflict with Higgs precision data unless the Higgs-to-Higgs-pair branching ratio is below about 0.1--0.3~\cite{Bernon:2014vta}, 
depending on the model.\footnote{A large survey of exotic Higgs decays is available in \cite{Curtin:2013fra}.}   

Only by tuning the model parameters so that the SM-like Higgs has very small coupling to a pair of lighter Higgs 
bosons can such a small branching ratio be achieved.  Nonetheless, this is a parameter space window that 
cannot yet be excluded and that has many interesting special features, including rather large predicted cross 
sections for direct production of the light Higgs boson(s) --- cross sections that might even be testable using 
the existing LHC 8 TeV data. The goal of this paper is to delineate these scenarios and their special properties.  

We note that these scenarios are not achievable in the MSSM because of the strong interrelations of the Higgs potential parameters required by supersymmetry; a light $\ha$ is simply not consistent within the MSSM when the $\hl$ has mass $125\gev$ (unless the Higgs sector is CP-violating). MSSM scenarios in which the $\hh$ has mass of $125\gev$ and $\mha,\mhl$ are below $\mhh$ have been constructed \cite{Carena:2013qia}, but those to date do not have $\mha,\mhl<125/2\gev$. 
In the NMSSM,  scenarios with a light $a_1$ and/or $h_1$ are possible
in light of the current data
\cite{Cerdeno:2013cz,Cao:2013gba,Bomark:2014gya,Huang:2013ima} but are
not the subject of this paper --- they typically imply small cross
sections for production of the light Higgs boson.\footnote{NMSSM
scenarios with a light $a_1$ and/or $h_1$ that appears in the decay of
a SM-like Higgs (\eg\ $h_2\to a_1a_1$, where $h_2$ is SM-like) have a
long history, the original paper being \cite{Gunion:1996fb}.}
%

The key consideration for this study is the magnitude of the coupling of the SM-like Higgs to a pair of the other Higgs bosons.  We employ the formulae found in \cite{Gunion:2002zf} extensively. There one finds the following results.
\bea
g\ls{\hl\ha\ha} &=&
   {-v}\bigl[\lamT\sbma-\lamU\cbma\bigr]\,, \label{ghaa}\\
g\ls{\hh\ha\ha} &=&  
   {-v}\bigl[\lamT\cbma+\lamU\sbma\bigr]\,, \label{gHaa}\\
g\ls{\hh\hl\hl} &=& {3v}\bigl[
   \lam\cbma\left(-\nicefrac{2}{3}+\sbmaii\right)-\lamhat\sbma(1-3\cbmaii)
    \nonumber \\
&&+(2\lamA-\lamT)\cbma\left(\nicefrac{1}{3}-\sbmaii\right)  
    -\lamU\cbmaii\sbma\bigr]
    \label{gHhh}
\eea
where
\bea
\lamT &=&
\quarter\stwob^2(\lam_1+\lam_2)+\lamtil(\sb^4+\cb^4)-2\lam_5-s_{2\beta}
c_{2\beta}(\lam_6-\lam_7)\,, \label{lamtdef}\\[5pt]
\lamU &=&\half s_{2\beta}(\sb^2\lam_1-\cb^2\lam_2+
c_{2\beta}\lamtil)-\lam_6\sb\sthreeb -\lam_7\cb\cthreeb\,.
\label{lamudef}
\eea
In the above, $\tanb=v_2/v_1$ is the ratio of the vevs of the two Higgs doublets, $v\equiv \sqrt{v_1^2+v_2^2}=246\gev$ and $\alpha$ is the mixing angle required to diagonalize the CP-even mass-squared matrix (see \cite{Gunion:2002zf} for details). 

In terms of $\gyxx$, where $Y$ is the SM-like Higgs and $X$ is the $A$ for $Y=h$ and either the $A$ or $h$ for $Y=H$,  we find
\beq
  R(XX)\equiv {\Gamma(Y \to XX) \over \Gamma (Y \to bb)} 
   = {1 \over 12} \left( {\gyxx v \over m_Y m_b} \right)^2 {\beta (m_X) \over \beta^3 (m_b)}
\eeq
with  $\beta(m_X)=\sqrt{1-4m_X^2/m_Y^2}$.  
Taking $m_Y=125\gev$ and assuming purely SM-like couplings for $Y$, one finds that $R(XX)\lesssim{5 \over 6}$ (equivalent to $\br(Y\to XX) < 0.3$) requires $|\gyxx|\lsim 17\gev$ for $m_X=62\gev$, which goes down to $|\gyxx|\lsim 6\gev$ for $m_X\simeq10-40\gev$. 
We will see that such a small $\gyxx$ is a very strong constraint --- without parameter tuning $|\gyxx|$ 
is most naturally of the order of a TeV. 

In the following, we consider $Y=h$ in Section~II and $Y=H$ in Section~III. 
We begin each of these sections by discussing the special parameter choices required in order to avoid 
too large Higgs-to-Higgs-pair branching ratio(s) for the $125\gev$ state and then proceed to the associated 
phenomenology. 
Our procedure for exploring the 2HDM parameter space is the same as in \cite{Dumont:2014wha,Dumont:2014kna}. 
All points that are retained obey the constraints from stability, unitarity and perturbativity (SUP), 
electroweak precision tests (STU), LEP searches, as well as the limits imposed by non-observation 
at the LHC of any Higgs bosons other than the SM-like one at $125\gev$. 
Regarding constraints from the Higgs signal strength measurements at $125\gev$, for each of the observed 
Higgs decay modes ($\gamma\gamma$, $WW^{(*)}$, $ZZ^{(*)}$, $b\bar b$, $\tau\bar\tau$) 
we require agreement at the 95\% CL with the ATLAS+CMS combined signal strength ellipse 
in the (ggF+ttH) and (VBF+VH) plane, as explained in \cite{Dumont:2014wha}. These signal strength ellipses 
have been determined from a fit with {\tt Lilith 1.0.1}~\cite{Bernon:2014vta,lilith}, 
including the lastest experimental results as of October, 2014. 
In the plots below we consider only scan points that pass all these constraints.\footnote{In \cite{Dumont:2014wha,Dumont:2014kna}, we also required that the ``feed down'' of heavier Higgs states to the signal at $125\gev$ be not too large. In the scenarios investigated in this paper, such feed down processes are irrelevant.}

\section{\boldmath The  $\mhl\sim 125\gev$ case}

Using Eqs.~(\ref{ghaa}), (\ref{lamtdef}), (\ref{lamudef}) and the relationships of the $\lam_i$ to the physical Higgs masses and the Higgs mixing parameter, $\mot^2$, in the scalar potential (see \cite{Gunion:2002zf}), one finds the following result for the $hAA$ coupling:
\bea
\label{ghaaform}
  g\ls{\hl\ha\ha} &=&  {1 \over 2 v}\left[\left(2\mha^2-\mhl^2\right) {\cos (\alpha-3 \beta) \over \sin2\beta} + 
  \left(8 \mot^2-\sin 2\beta \left( 2 \mha^2+3 \mhl^2\right)\right) {\cbpa \over \sin^2 2\beta}\right]
\eea
Let us begin by taking 
the SM limit,  $\sbma=1$, in the formula above:
\beq
g_{hAA}= - {2\mha^2+\mhl^2-2\mhat^2 \over v}
\label{ghaaform2}
\eeq
where $\mhat^2=\mot^2 \sec\beta \csc\beta$ and $\mot^2$ can be positive or negative.  Given that $|g_{hAA}|$ must be very small to have small $\br(h\to AA)$, we see that in this limit $\mhl^2\sim -2\mha^2+2\mhat^2$ is required.\footnote{Without this cancellation, when the $\ghaa$ coupling is large, one may still suppress the $h \to AA$ decay by minimizing its phase space; however, this is not the case of interest in this study.}  
While there is no symmetry that motivates this particular choice, it can certainly be satisfied for appropriately modest $\mhat^2$ and we find many allowed points of this nature.  

\begin{figure}[t]
\begin{center}
\includegraphics[width=0.6\textwidth]{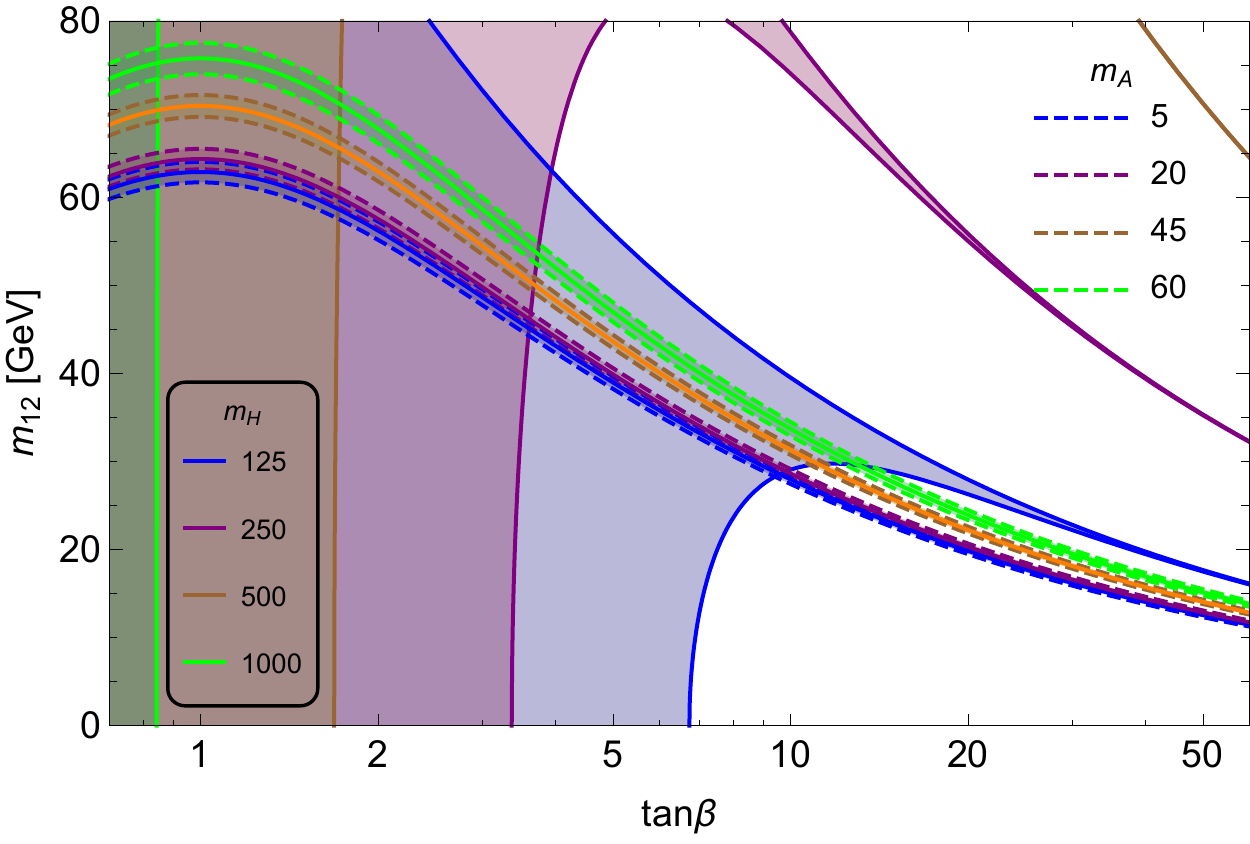}
\end{center}\vspace*{-5mm}
\caption{For  $\sbma=1$, we show the regions of $\mot$ vs.\ $\tanb$ parameter space consistent with perturbativity for various $\mhh$ values (see in-figure color code in lower-left corner).  Also shown are the narrow regions for which  $\br(h\to AA)<0.3$, assuming $h$ is the SM-like Higgs at 125~GeV with a total decay width of 4.07 MeV, for the indicated values of $\mha$ shown in the upper-right corner.  The figure applies to both the \typei\ and \typeii\ 2HDM. The perturbatively acceptable region also extends to $\mot^2<0$, but this region is not plotted since \Eq{ghaaform2} 
would give large $|\ghaa|$ and, therefore, large $\brhaa$ if $\mot^2$ were negative. 
}
\label{pertfigh}
\end{figure}

The interrelations of the parameters in this region are illustrated in Fig.~\ref{pertfigh}. The figure shows the combined impact of perturbativity and the requirement of small $\br(h\to AA)$.  
The large solid filled regions are those allowed by perturbativity for various different values of $\mhh$ (as indicated by the color code in the lower-left corner of the plot). The regions surrounded by dashed lines are those consistent with $\br(h\to AA)\leq 0.3$, with the central solid line corresponding to $\br(h\to AA)=0$ (or equivalently $\ghaa$=0 ), for the various $\mha$ values coded as shown in the upper-right corner of the plot. 
We see that the higher the  value of $\mhh$, the smaller the $\tanb$  that is required by perturbativity.  
Imposing both perturbativity and $\br(h\to AA)\leq 0.3$ strongly constrains $\mot$ within the allowed $\tanb$ range (note: $\mot\equiv\sgn(\mot^2)\sqrt {|\mot^2|}$). Roughly, $\mot\approx 30-100\gev$ and $\tanb<15$ are the interesting ranges to scan over for this solution. 

Deviating from the strict SM limit, there is also another parameter region that gives small $|\ghaa|$ 
through a cancellation between the first and second terms in  \Eq{ghaaform} 
(or, equivalently, between the $\mot^2$ and non-$\mot^2$ terms in this equation).  
This can be achieved when $\sbpa$ is close to one and allows also for larger $\mot^2$. 
As described in \cite{Ferreira:2014naa}, $\sbpa\sim 1$ can be consistent with the $h$ being SM-like so long as $\tanb$ is not too small. In particular, one finds in this limit 
\beq
\cv=\sbma\to {\tan^2\beta -1\over \tan^2\beta+1}\,,
\eeq
where $\cv$ is the magnitude of the $hVV$ coupling relative to the SM value.  
One obtains $\cv\gsim 0.95$ once $\tanb\gsim 6$, \ie\ sufficiently close to unity for consistency with 
Higgs data from the LHC.  
Note, however, that one cannot actually use exactly $\sbpa=1$.  This is because if both $\sbma\to1$ and $\sbpa\to1$, 
then $\beta\to\pi/2$ and $\alpha\to0$, for which $\ghaa$ becomes too large. Indeed, in the limit of $\sbpa=1$, 
we obtain
\beq
\ghaa=  {2\mha^2-\mhl^2 \over v}\, \cos2\beta\,,
\eeq
which is too large given that $\cos2\beta\sim -1$ for $\tanb\gsim 6$. 

\begin{figure}[t]
\begin{center}
\includegraphics[width=0.5\textwidth]{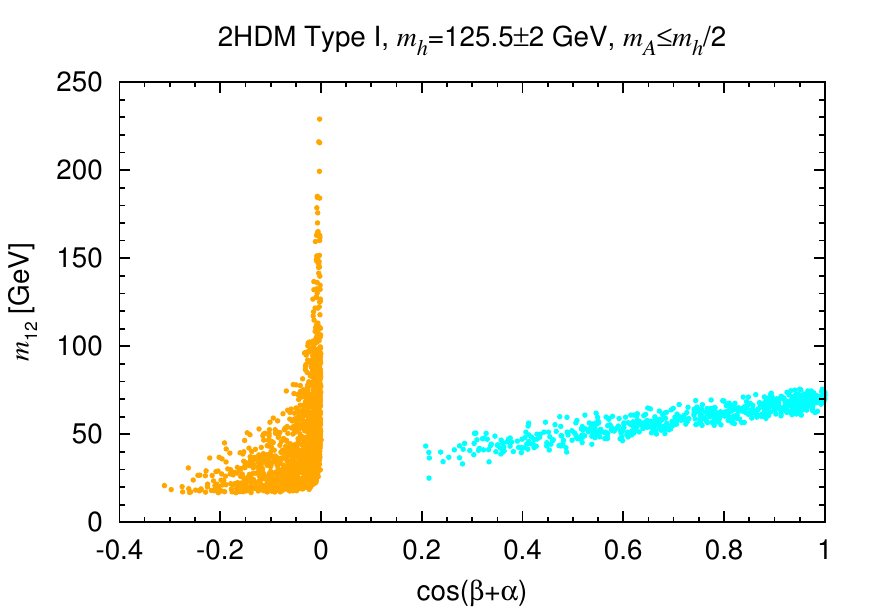}\hskip-.1in
\includegraphics[width=0.5\textwidth]{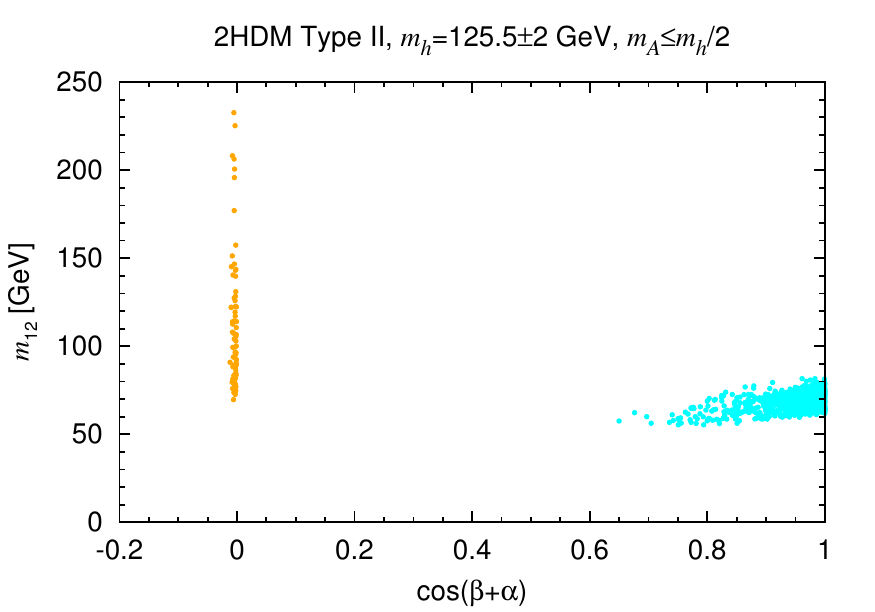}
\end{center}\vspace*{-5mm}
\caption{Phenomenologically viable points with $m_A\leq m_h/2$ in the $\mot$ vs.\ $\cbpa$ plane,  for 
2HDM \typei\ (left) and \typeii\ (right). The cyan points have $\sbma\sim 1$, $\cbma>0$ and modest 
$\mot$, as for the  $\sbma=1$ allowed region seen in Fig.~\ref{pertfigh}, while the orange points have 
$\sbpa\sim 1$, small $\cbpa<0$ and $\tanb>5$.}
\label{insight2}
\end{figure}

\begin{figure}[t]
\begin{center}
\includegraphics[width=0.5\textwidth]{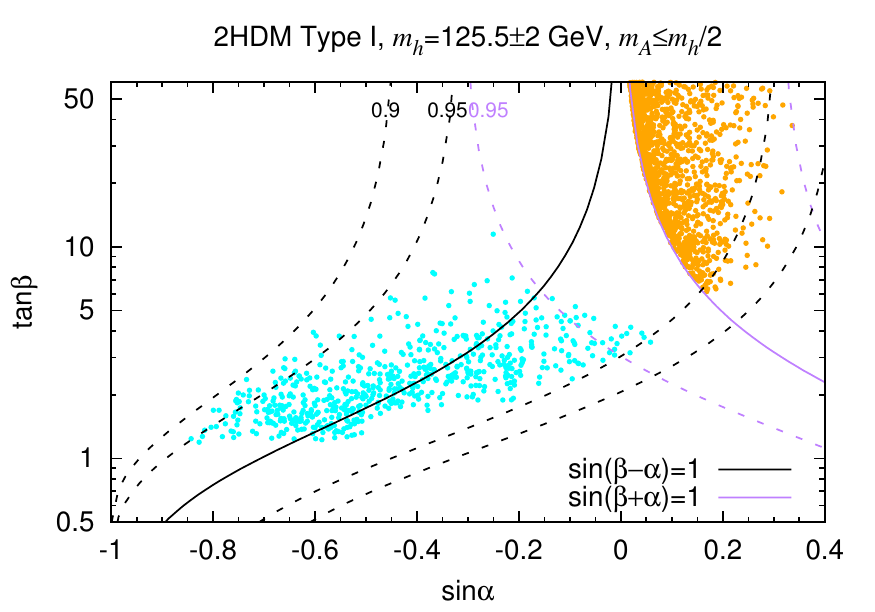}\hskip -.1in
\includegraphics[width=0.5\textwidth]{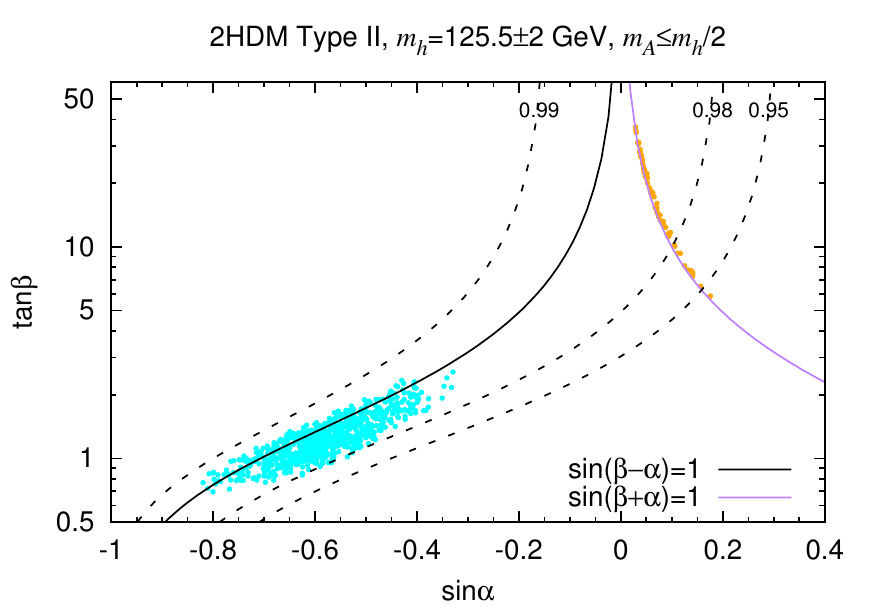} 
\end{center}\vspace*{-5mm}
\caption{Same as Fig.~\ref{insight2} but in the $\tanb$ vs.\ $\sina$ plane. The solid black and purple lines indicate 
$\sbma=1$ and $\sbpa=1$, respectively. The dashed black (purple) lines are iso-contours of  values of 
$\sbma$ ($\sbpa$) as indicated on the plots.}
\label{haafig1}
\end{figure}

An overall view of the allowed low-$m_A$ points in $\mot$ vs. $\cbpa$ space for the \typei\ and \typeii\ 2HDMs is provided by \Fig{insight2}, and in the $\tanb$ vs.\ $\sina$ plane in \Fig{haafig1}. The cyan points have $\sbma\sim 1$, $\cbpa>0$ and modest $\mot$, as for the  $\sbma=1$ allowed region seen in Fig.~\ref{pertfigh}, while the orange points are those with $\sbpa\sim 1$, small $\cbpa<0$, $\tanb>5$ and $m_{12}>0$. (The opposite case with $\mot<0$ and $\cbpa>0$ could also lead to the necessary cancellations in  \Eq{ghaaform} but turns out to be excluded by the 125~GeV Higgs signal constraints.) 
In \Fig{haafig1}, points to the right of the $\sbpa=1$ curve have $\cbpa<0$ and those to the left have $\cbpa>0$.  The requirement of small $\ghaa$ (coupled with $\mot>0$) thus creates a very sharp boundary between acceptable vs.\ non-acceptable parameter points.  One should also note that the $\sbpa\sim 1$ points mostly (entirely) have $\sina>0$  for \typei\ (\typeii).  Consequently, in the \typeii\ model the orange points correspond to the ``wrong-sign'' Yukawa coupling $\cd^{\hl}\sim -1$ \cite{Ferreira:2014dya}, whereas the cyan points have $\cd^{\hl}>0$.

\begin{figure}[t]
\vspace*{-10mm}
\begin{center}
\hspace*{-16mm}\includegraphics[width=0.7\textwidth]{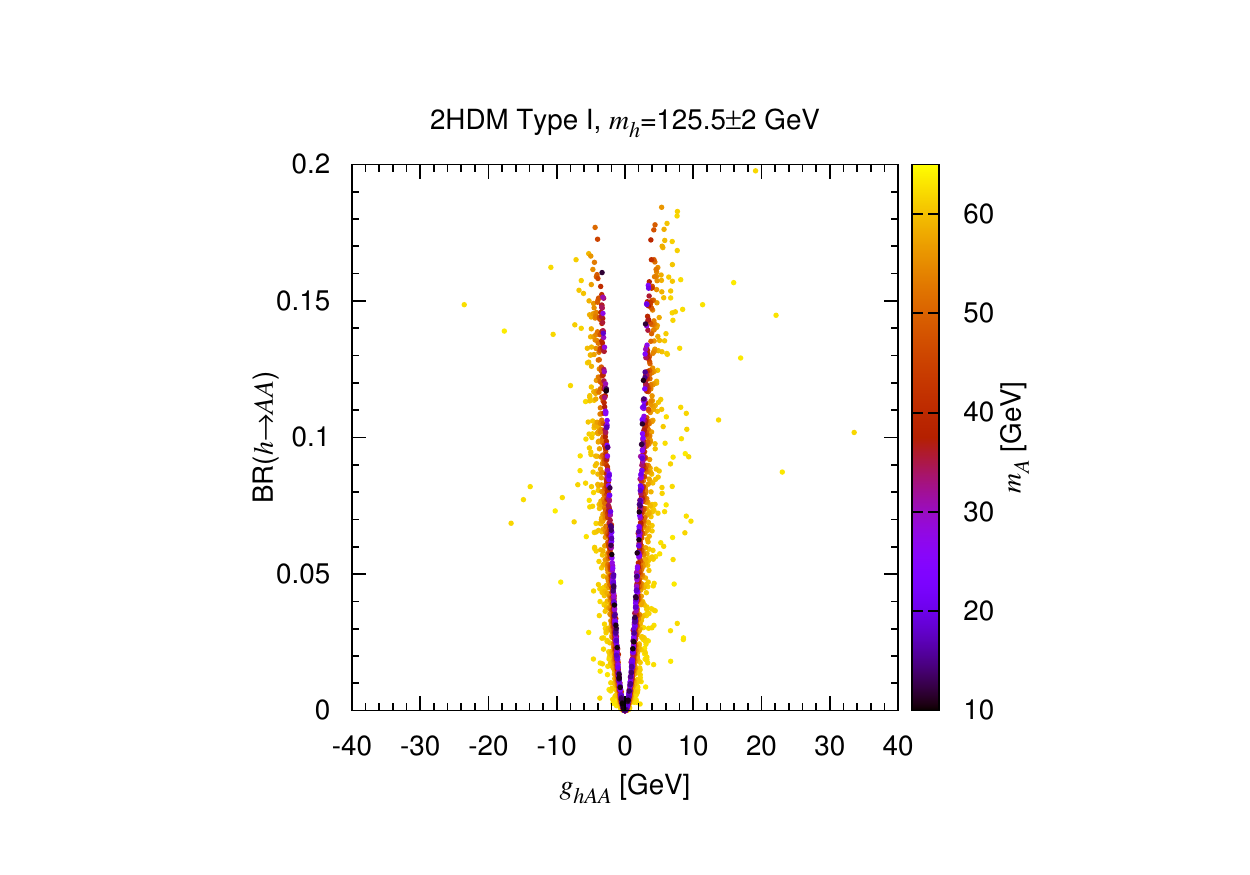}\hspace*{-33mm}\includegraphics[width=0.7\textwidth]{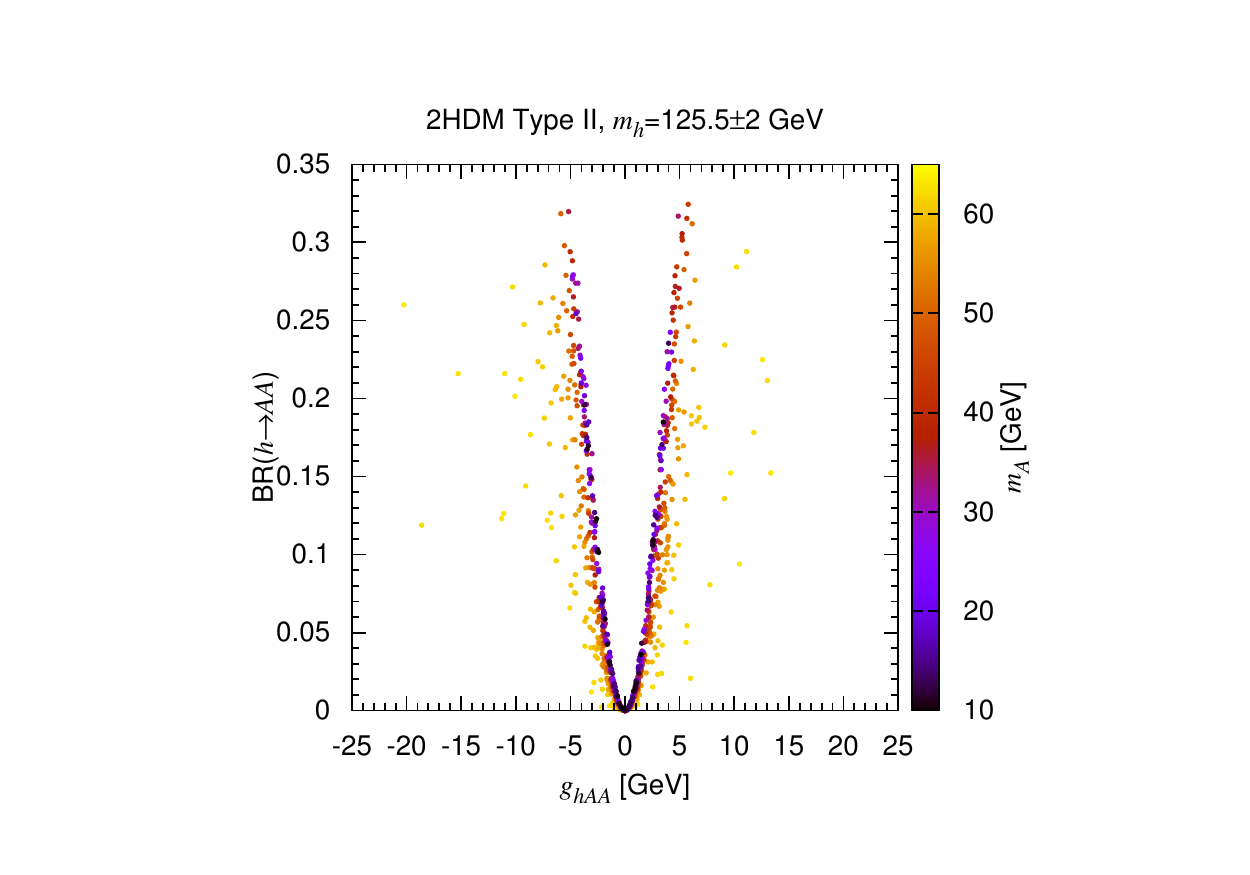} 
\end{center}\vspace*{-10mm}
\caption{Allowed points in the BR$(h\to AA)$ vs.\ $\ghaa$ plane, on the left for Type~I, on the right for Type~II. The value of $m_A$ is colour-coded as indicated by the scales on the right of the plots.}
\label{ghAA-BRhAA}
\end{figure}

For completeness we show in Fig.~\ref{ghAA-BRhAA} the explicit values of BR$(h\to AA)$ vs.\ $\ghaa$ 
for the allowed points. We see that $\ghaa$ is indeed tightly constrained to small values of the order of 5~GeV. 
Note that the allowed range for BR$(h\to AA)$ is different for Type~I and Type~II models because of the different structure of the $h$ couplings to fermions. As an aside we also note that agreement with the individual 95\% CL signal strength ellipses allows for slightly higher BR$(h\to AA)$ than a global fit, which would 
restrict BR$(h\to AA)<0.16$ (0.26) in Type~I (Type~II) at 95\% CL. 

\begin{figure}[t]
\begin{center}
\includegraphics[width=0.5\textwidth]{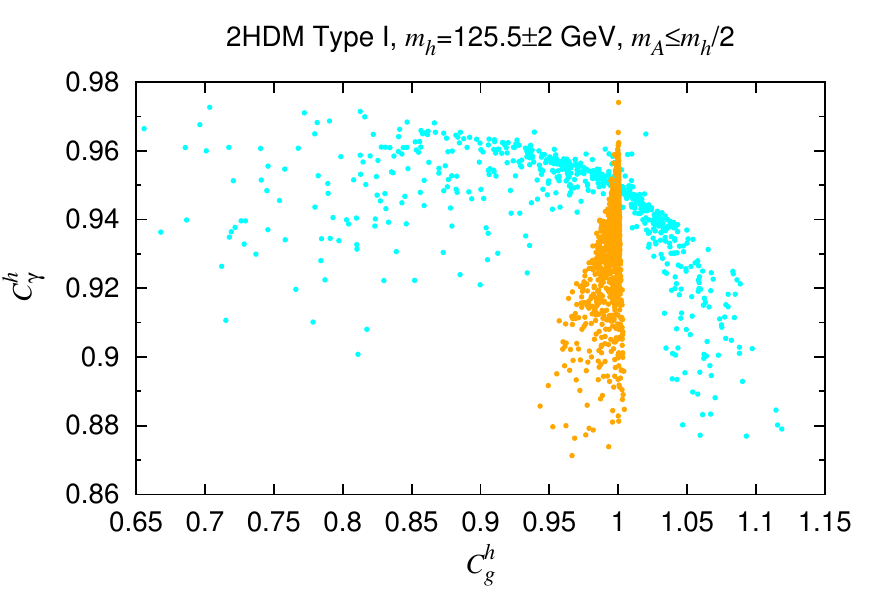}\hskip-.1in
\includegraphics[width=0.5\textwidth]{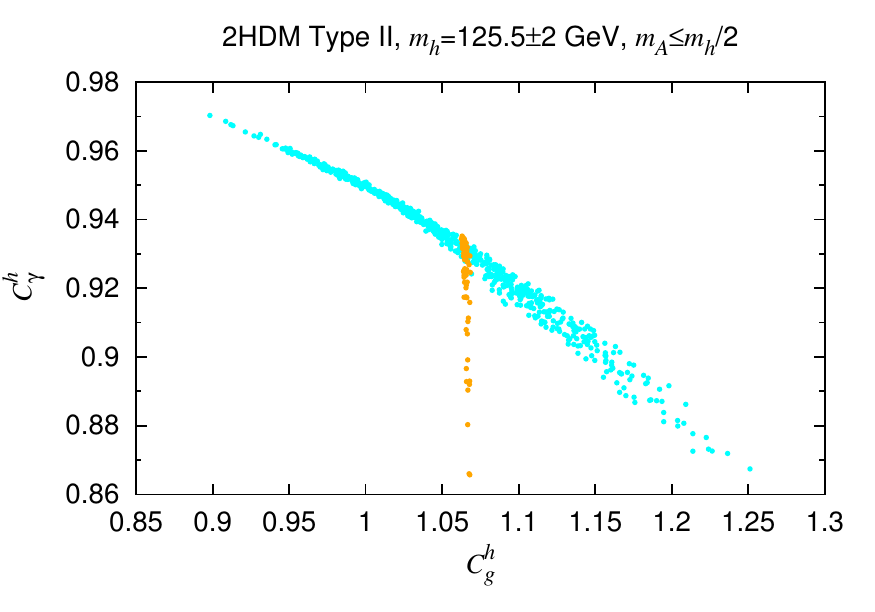} 
\end{center}\vspace*{-5mm}
\caption{As Fig.~\ref{insight2} but for $\cp^h$ vs.\ $\cg^h$. }
\label{cgvscgam}
\end{figure}

\begin{figure}[t]
\begin{center}
\includegraphics[width=0.5\textwidth]{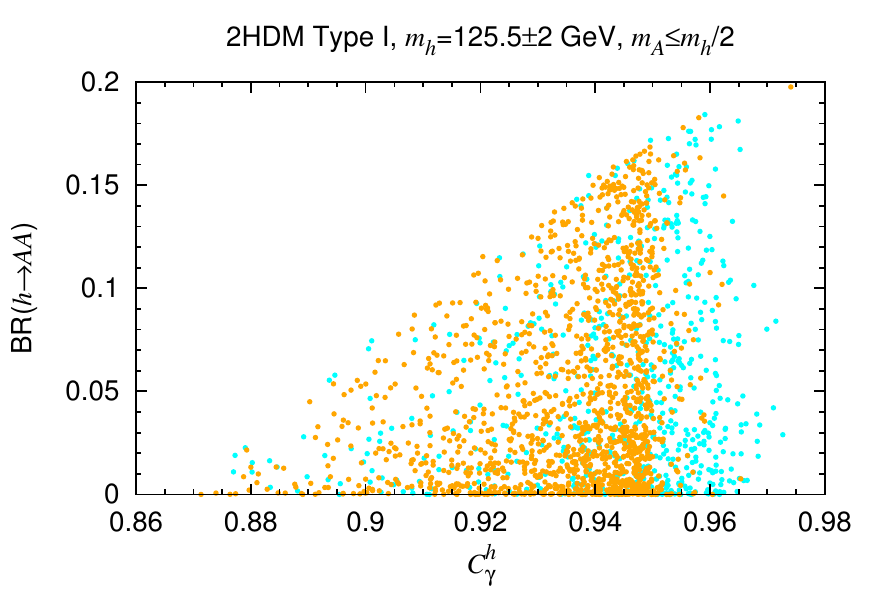}\hskip-.1in
\includegraphics[width=0.5\textwidth]{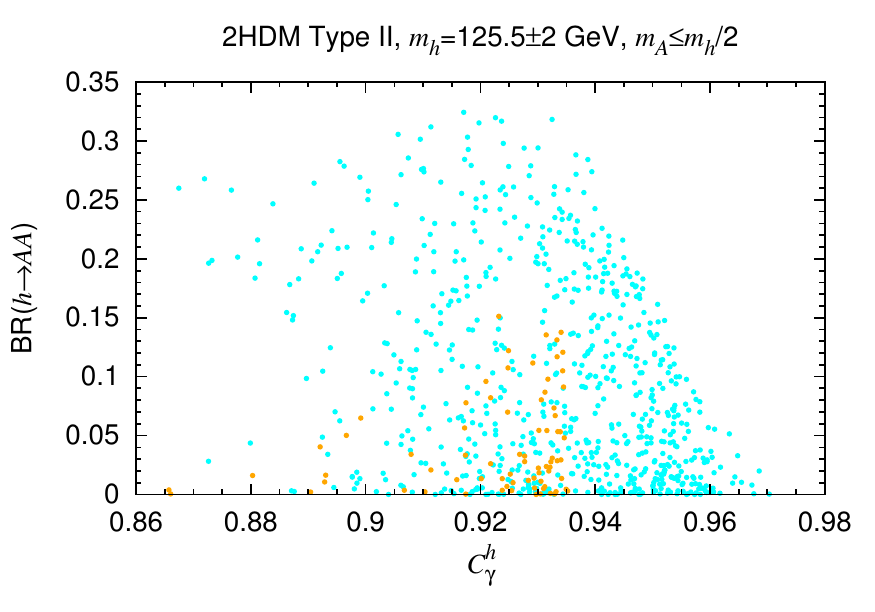} 
\end{center}\vspace*{-5mm}
\caption{As Fig.~\ref{insight2} but for BR$(h\to AA)$ vs.\ $\cp^h$.}
\label{brhaavscgam}
\end{figure}

Having understood the constraints on this scenario, we now pursue the implications for LHC phenomenology. 
In \Fig{cgvscgam} we plot the reduced couplings (relative to their SM values) of $h$ to gluons and to photons, 
$\cg^h$ vs.\ $\cp^h$. 
The suppressed values of $\cp^h$ come from the negative contribution of the charged Higgs to the $h\gam\gam$ one-loop coupling. In the limit of $\sbma=1$, 
\beq 
  g_{hH^\pm H^\pm}=g_{hAA}-(\lam_5-\lam_4)v=g_{hAA}-2(m^2_{H^\pm}-m^2_A)/v \,. 
\eeq  
The first term, $g_{hAA}$, has to be small as discussed above and the second term is always negative 
because $m_{H^\pm} \gsim 90\gev ~(300\gev)$ in Type I (Type II). 
The relation between $\cp^h$ and BR$(h\to AA)$ is shown in Fig.~\ref{brhaavscgam}. 
While in Type~I the allowed value of BR$(h\to AA)$ increases with $\cp^h$, 
the trend is the opposite in Type~II though with a much less pronounced correlation. 

\begin{figure}[t]
\begin{center}
\includegraphics[width=0.5\textwidth]{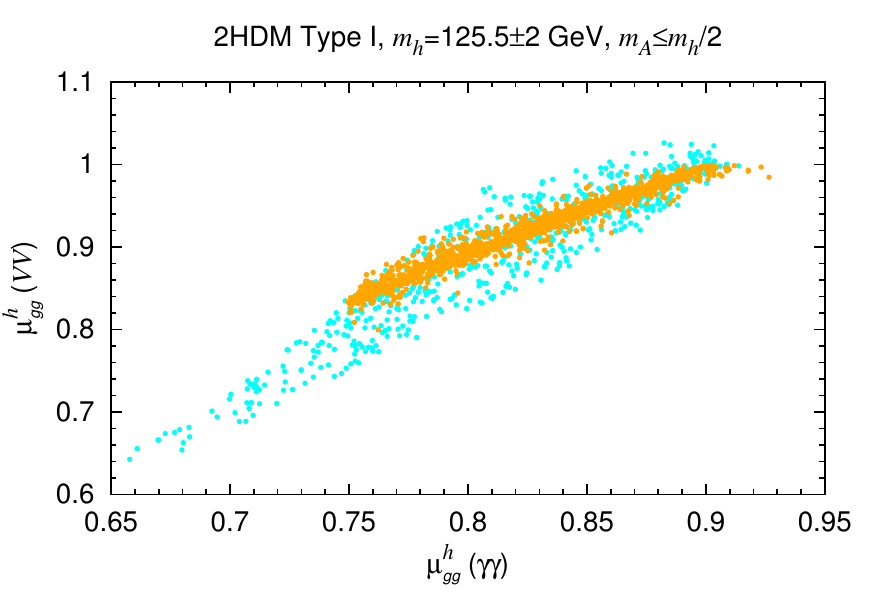}\hskip-.1in
\includegraphics[width=0.5\textwidth]{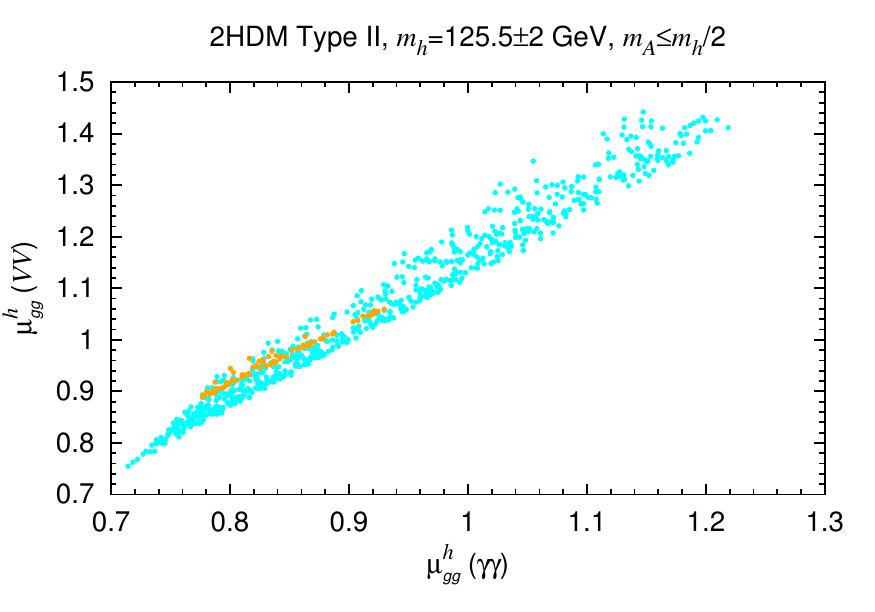} 
\end{center}\vspace*{-5mm}
\caption{Signal strengths $\rgghzz$ vs.\ $\rgghgamgam$ for the \typei\ and \typeii\ models.  
The orange points are, as for previous plots, the points with $\sbpa\sim 1$.
}
\label{muratios}
\end{figure}

To illustrate the impact on observables, we plot in \Fig{muratios} the signal strengths $\mu$ (relative to the SM) for $gg\to h\to VV$ ($V=W,Z$) versus $gg\to h\to \gam\gam$, denoted as $\rgghzz$ vs.\ $\rgghgamgam$.  
Our first observation is that $\rgghgamgam$ is suppressed for all points in \typei\ as well as for the orange points in \typeii. 
The deviations from the SM predictions of unity are of course consistent with current data, since this was a requirement of the scan, but it is obvious that future higher precision measurements will strongly constrain these scenarios.  
Remarkably --- and in contrast to the case when $m_A>m_h/2$ --- it is impossible to simultaneously achieve 
$\rgghgamgam=1$ and $\rgghvv=1$ in either \typei\ or \typeii\ when $m_A\leq m_h/2$. 
(See Fig.~2 of \cite{Dumont:2014kna} for comparison with the general case.) 
Thus, this scenario will be excluded should the Higgs observations converge sufficiently close to the SM expectations. 

\begin{figure}[t]
\begin{center}
\includegraphics[width=0.5\textwidth]{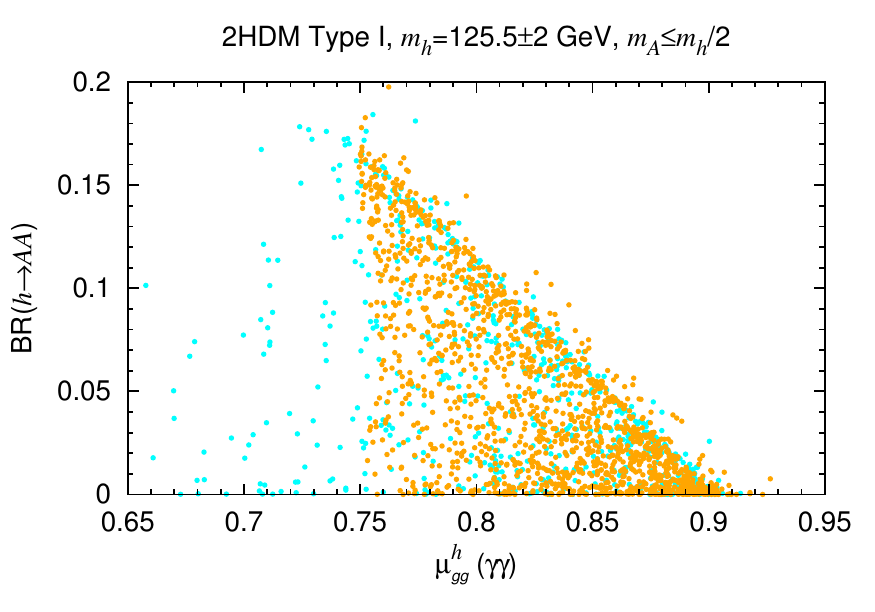}\hskip-.1in
\includegraphics[width=0.5\textwidth]{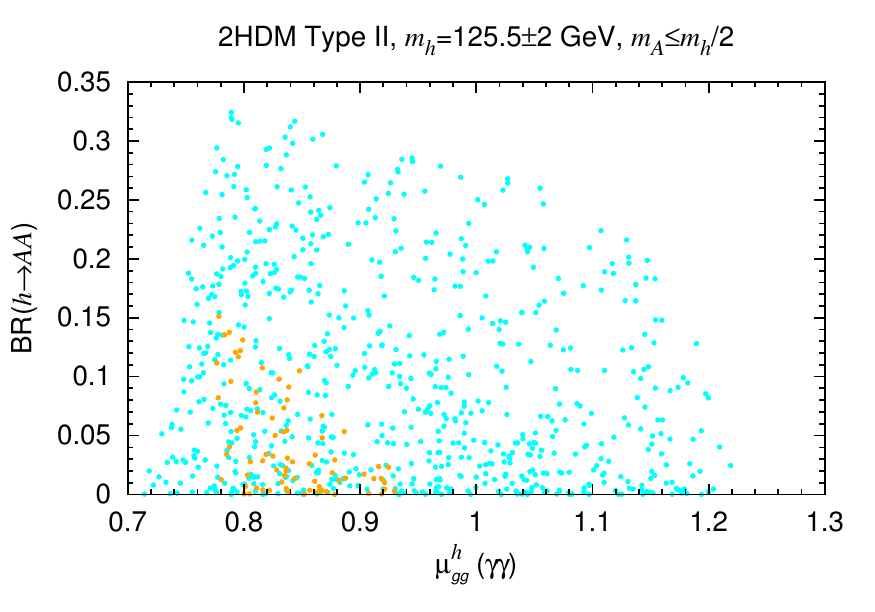} 
\end{center}\vspace*{-5mm}
\caption{BR$(h\to AA)$ vs.\ $\rgghgamgam$ for the \typei\ and \typeii\ models.}
\label{brhaavsmu}
\end{figure}

Figure~\ref{brhaavsmu} shows BR$(h\to AA)$ vs.\ signal strength $\rgghgamgam$. From the left plot we can directly see that in Type~I a precise measurement of this signal strength gives an upper bound on the allowed $h\to AA$ branching ratio. If $\rgghgamgam$ is measured to be within 10\% of unity, this  means  BR$(h\to AA)\lsim 0.01$. Conversely, 
a measurement of $\rgghgamgam\simeq 1$ combined with detection of $h\to AA$ decays implies that the Type~II model is strongly preferred and that the wrong-sign Yukawa solution is excluded. 

\begin{figure}[t]
\begin{center}
\includegraphics[width=0.5\textwidth]{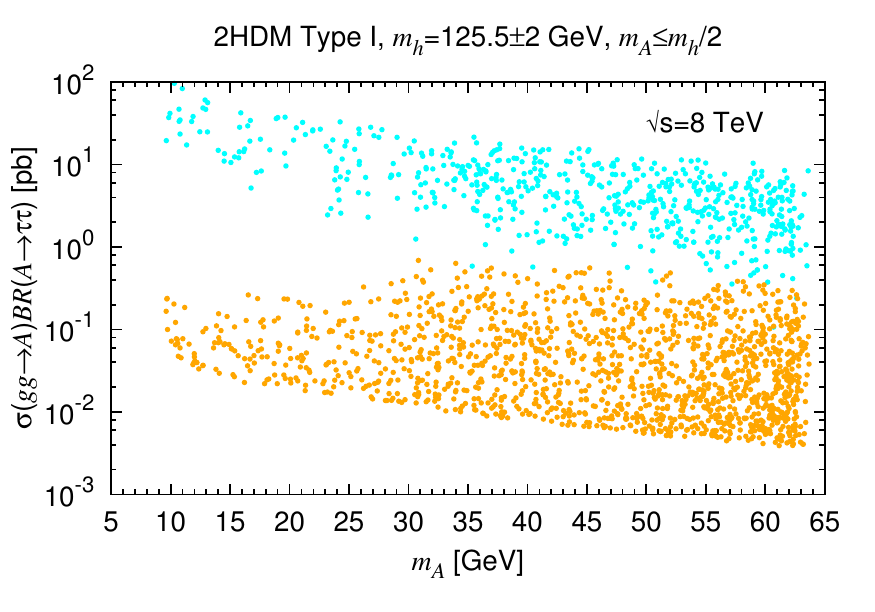}\hskip-.1in
\includegraphics[width=0.5\textwidth]{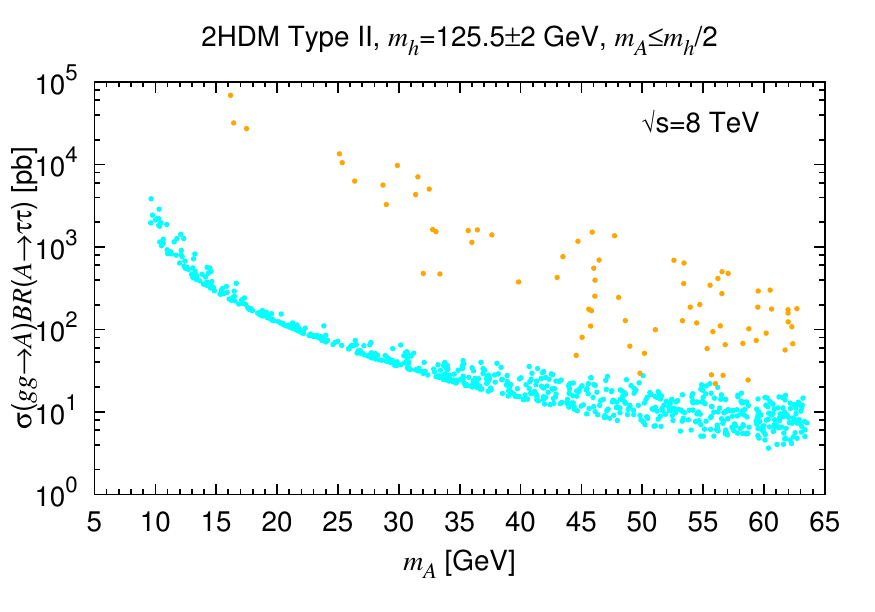}\\
\includegraphics[width=0.5\textwidth]{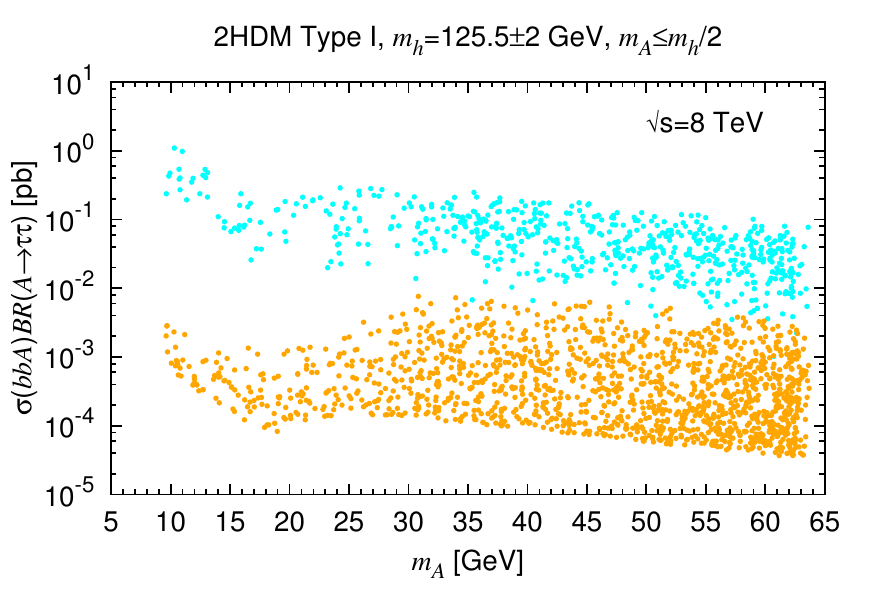}\hskip-.1in
\includegraphics[width=0.5\textwidth]{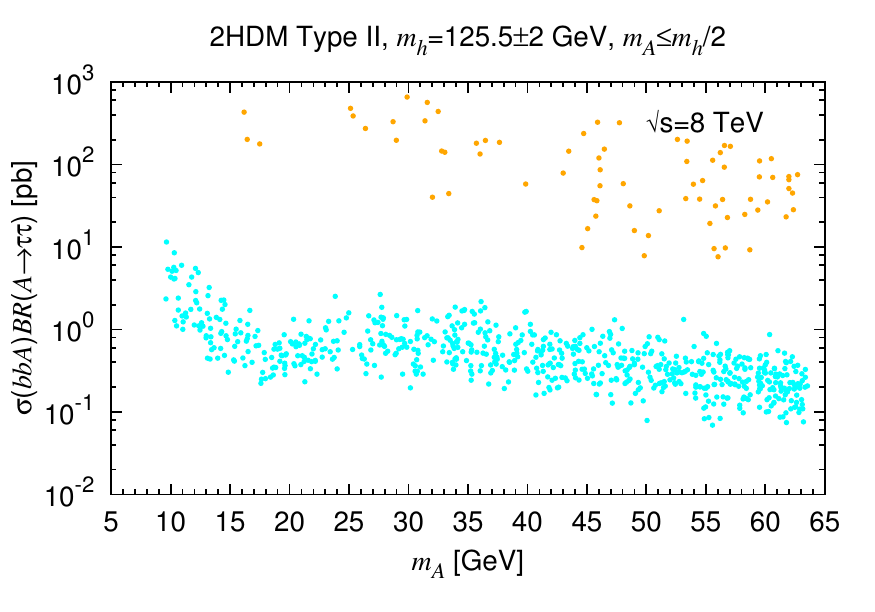}
\end{center}\vspace*{-5mm}
\caption{Cross sections at $\sqrt{s}=8$~TeV for light $\ha$ production from $gg$ fusion (top row) and $b\anti b$ associated production (bottom row) in the $\tau\tau$  final state. The cross sections for the $\mu\mu$ final state have exactly the same form but are two orders of magnitude lower. Same colour scheme as in the previous figures.}
\label{hxsecs}
\end{figure}

Let us now turn to the question of the size of the cross sections for $\ha$ production with decays to the potentially 
observable $\tau\tau$ and $\mu\mu$ final states.  Figure~\ref{hxsecs} shows the $gg$ fusion and $b\anti b$ associated 
production cross sections at $\sqrt{s}=8$~TeV times BR$(A\to\tau\tau)$. As can be seen, the $A\to\tau\tau$ signal 
can have quite substantial cross sections over the whole mass range considered. 
The cross sections for the $A\to\mu\mu$ signal have exactly the same shape but are about a factor of 100 lower. 
For reference, naive estimates suggest that, before cuts and efficiencies, for the existing 8~TeV dataset 
with integrated luminosity of $L\simeq20\fbi$,  
a cross section of order $10\pb$ ($200,000$ events) should be observable in the 
$\tau\tau$ final state while $0.1\pb$ (2000 events) should be observable in the $\mu\mu$ final state, especially 
in the case of $b\anti b$ associated production by using modest $p_T$ $b$-tagging. From \Fig{hxsecs}, we observe that these levels are reached in the case of \typeii\ for essentially the entire $\mha\leq \mhl/2$ region in the case of $gg$ fusion and for the orange points in the case of $b\anti b$ associated production.\footnote{Recall  from \Fig{haafig1} that the orange points can have high $\tanb$ 
while the cyan points have quite modest $\tanb$ values.  This implies that the $b\anti b$ coupling in the \typei\ (\typeii) 
model is suppressed (enhanced). As a result, the orange points have the smallest (largest) cross sections in the 
case of \typei\ (\typeii).} Indeed, the cross sections for the orange points are really very large and should produce readily observable peaks.  In the case of the \typei\ 2HDM, many of the cyan points have $gg$ fusion cross sections at the probably observable $10\pb$ ($0.1\pb$) level in the $\tau\tau$ ($\mu\mu$) final states, but the orange points have cross sections that are almost certainly too small for detection in the Run~1 data set. 

Analyses by ATLAS and CMS for such signals at low $\mha$ in the $\tau\tau$ channel have significant background from the $Z$ peak.  As a result, limits are currently only available for $\mha\gsim \mz$. We are unaware of any public results for the $\mu\mu$ final state in the low mass region, but the excellent mass resolution in this channel should make separation from the $Z$ peak straightforward.

Finally, we note that running at higher energies will not straightforwardly improve the sensitivity to the low $m_A$ region, as the cross sections at 13--14~TeV are barely a factor 2 larger than those at 8~TeV. Therefore, one will need to accumulate more statistics via higher total integrated luminosity.

\section{\boldmath The $\mhh\sim125\gev$ case}

We first note that for $\mhh\sim125\gev$ in the \typeii\ model, $B$-physics constraints require 
$\mha\gsim 200\gev$. We therefore only consider the case of $\mhl<\mhh/2$ for \typeii.  In contrast, 
in the \typei\ model either $\mha$ or $\mhl$ can be $<\mhh/2$, but LEP limits imply that not both can be light 
simultaneously. This latter follows from the fact that  the  $\hh VV$ coupling and the $Z\hl\ha$ coupling 
are both proportional to $\cbma$.  Thus, for a SM-like $\hh$, \ie\ $|\cbma|\sim 1$ as required by signal strengths measurements, the  $Z\hl\ha$ coupling is near maximal 
and therefore the $Z^*\to \hl\ha$ cross section at LEP is too large, barring phase-space suppression.  

In practice we can therefore consider the $\hh\to \ha\ha$ and $\hh\to \hl\hl$ cases independently of one another.
With this in mind, we turn to the conditions for achieving small trilinear couplings in order to evade too large 
BR$(H\to AA)$ or BR$(H\to hh)$. Analogous to \Eq{ghaaform} we find 
\beq
\label{gHaaform}
   g\ls{\hh\ha\ha} =  {1 \over 2 v}\left[\left(2\mha^2-\mhh^2\right) {\sin (\alpha-3 \beta) \over \sin2\beta} 
   + \left(8 \mot^2-\sin 2\beta \left( 2 \mha^2+3 \mhh^2\right)\right) {\sbpa \over \sin^2 2\beta}\right] 
\eeq
and
\beq
\label{gHhhform}
   g\ls{\hh\hl\hl} =  -{1 \over v} \cbma \left[ {2\mot^2 \over \sin 2\beta}
   + \left( 2 \mhl^2+ \mhh^2 - {6 \mot^2 \over \sin 2\beta} \right) {\sin 2\alpha \over \sin 2\beta} \right] \,.
\eeq

\begin{figure}[t]
\begin{center}
\includegraphics[width=0.6\textwidth]{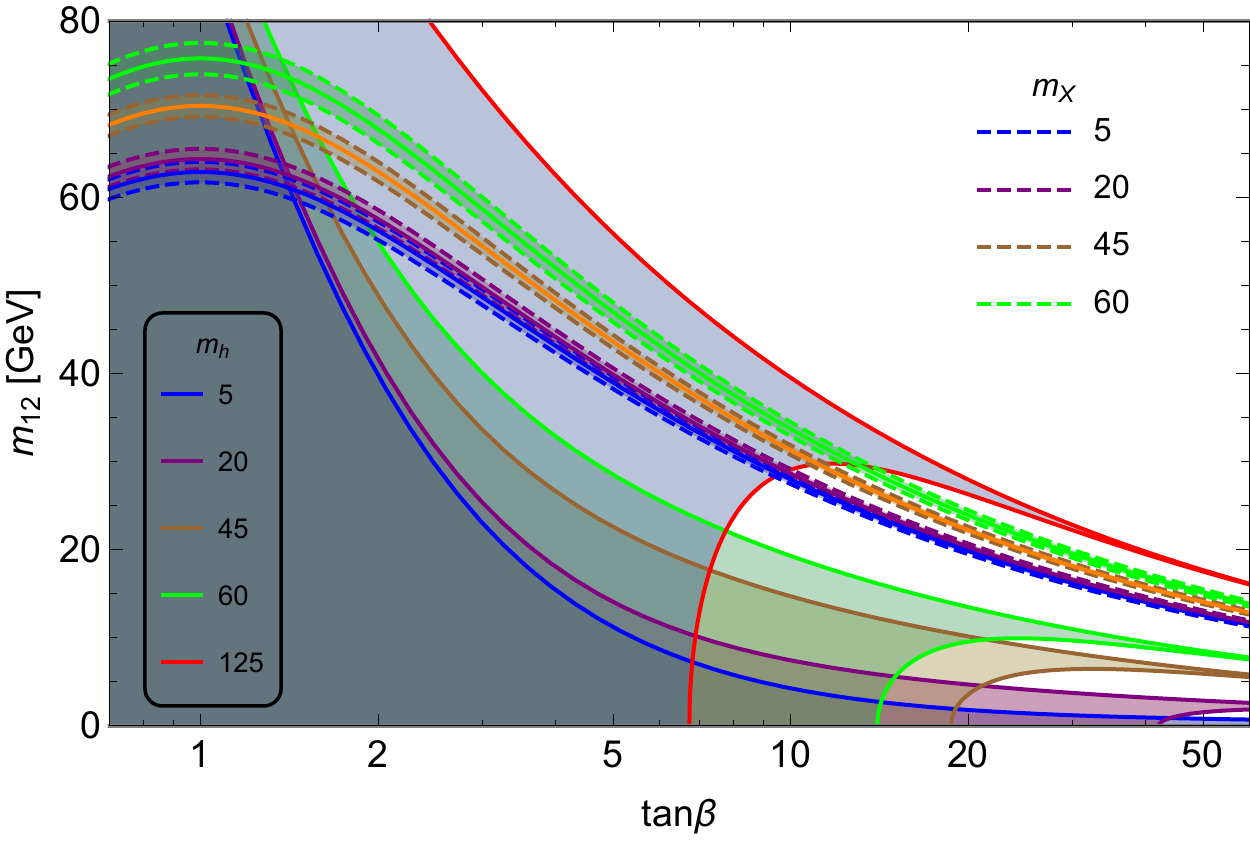}
\end{center}\vspace*{-5mm}
\caption{Constraints in the $\mot$ vs.\ $\tan\beta$ plane for the $H125$ case with $|\cbma|=1$. The shaded regions are those allowed by perturbativity for $\mhl$ values indicated in the lower-left corner of the plot. 
The narrow strips between the dashed lines have $\br(\hh\to XX)<0.3$  for $\mha<\mhh/2$ or $\mhl<\mhh/2$, respectively (the regions are the same for the two cases) with the colour code for the $X=h~\textrm{or}~A$ masses given in the upper-right corner of the plot. The solid line in the middle of the dashed ones shows $g\ls{HXX}=0$. 
}
\label{pertfigH}
\end{figure}

\bigskip
\noindent
As mentioned, for the $\hh$ to be SM-like, we should have $|\cbma|$ close to unity.  
One class of scenarios is easily understood by taking the strict limit of  $|\cbma|=1$, 
yielding
\beq
  g\ls{HXX}= - {2m_X^2+\mhh^2-2\mhat^2 \over v}\,, \qquad X=h,A\,. 
\label{gHxxform}
\eeq
Analogous to the $h125$ case, $\mhat^2=\mot^2 \sec\beta \csc\beta$ should be small and positive 
to achieve small enough $|g\ls{HXX}|$. 
The interplay of the requirements of perturbativity and of small $|g\ls{HXX}|$ is illustrated in  \Fig{pertfigH}. 
We see that for $m_h\le 60\gev$, small $\tan\beta$ below about 2 is required. 
(Note also that if both $h$ and $A$ were light, they should be very close in mass to suppress $\br(\hh\to \hl\hl,\ha\ha)$;  
this follows from the fact that the bands of $\br(\hh\to XX)<0.3$ are valid for both $X=h$ and $X=A$.)
For $0.5\,m_H<m_h<m_H$, \ie\ if only $A$ is light, there is a bit more freedom and $\tan\beta$ can go up to 10--15, 
tightly related however with $m_{12}$ for any given value of $m_A$. 
Figure~\ref{pertfigH} gives a somewhat idealized picture because the signal strength measurements at 
~125 GeV only require  $C_V\gsim 0.9$, and constraints from the oblique parameters STU actually 
forbid $|\cbma|$ being exactly 1;  nonetheless \Fig{pertfigH} serves as useful guidance for the parameter scan. 

As in the $h125$ case,  sufficiently small $|g\ls{HXX}|$ can also be achieved by resorting to cancellations 
between the various terms in \Eq{gHaaform} or \Eq{gHhhform}. In the $H125$ case, the $|\cbma|=1$ component shown in \Eq{gHxxform} is positive for larger  $\mot$ values than those shown in \Fig{pertfigH} and this component can be cancelled by the remaining term(s) for $\cbpa\sim1$.

\begin{figure}[t]
\begin{center}
\includegraphics[width=0.5\textwidth]{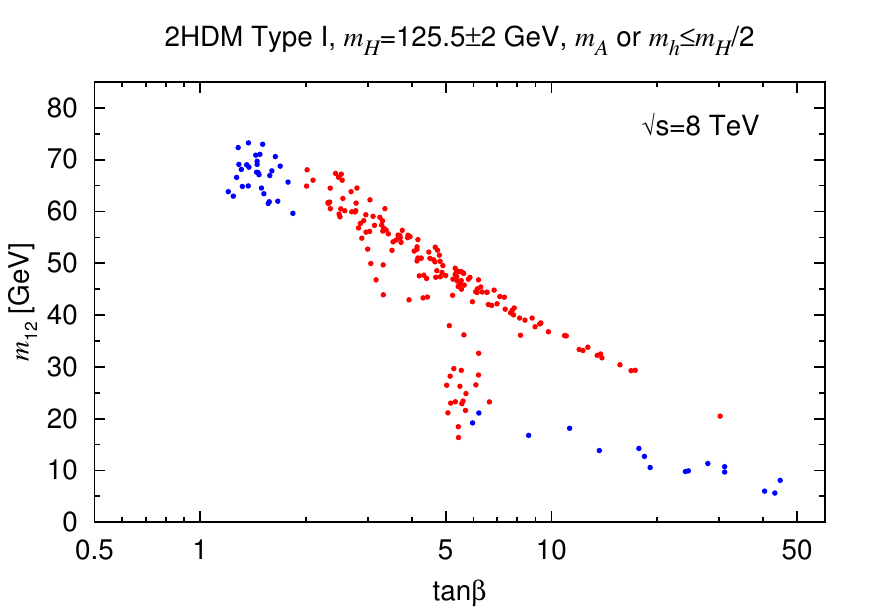}\hskip-.1in
\includegraphics[width=0.5\textwidth]{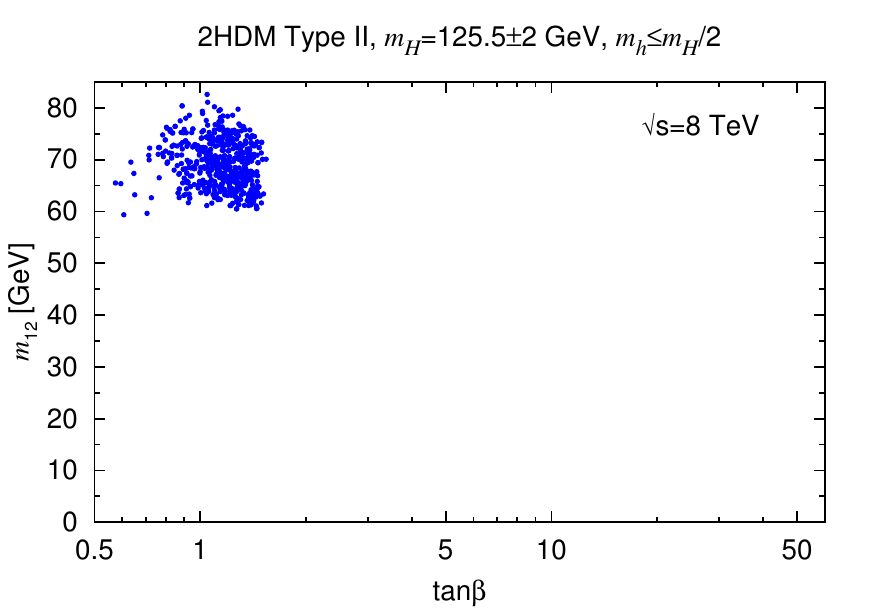}\\
\includegraphics[width=0.5\textwidth]{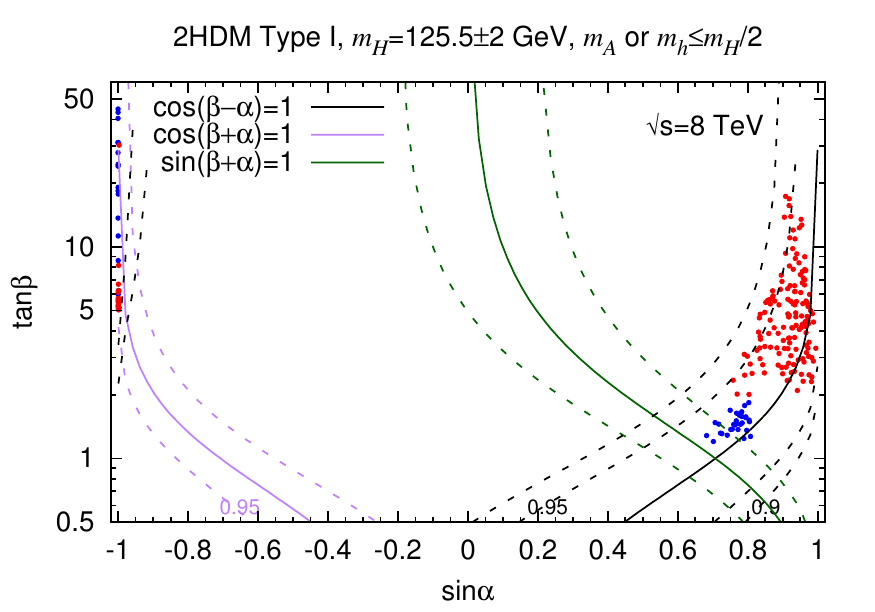}\hskip-.1in
\includegraphics[width=0.5\textwidth]{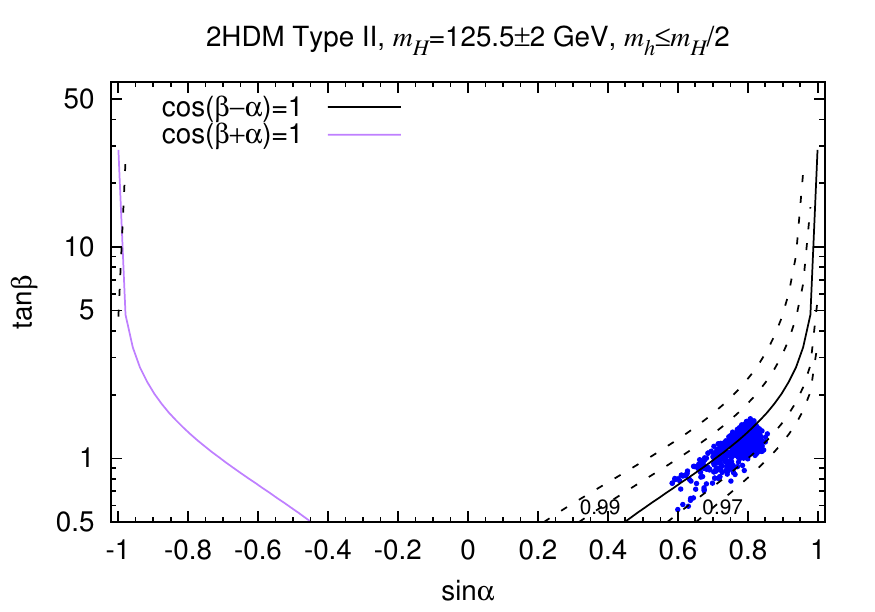}
\end{center}\vspace*{-5mm}
\caption{Phenomenologically viable scan points for the H125 scenario in the Type~I (left) and Type~II (right) models. 
The upper row shows the projection onto the $m_{12}$ vs.\ $\tanb$ plane for comparison with \Fig{pertfigH}. 
The lower row shows the $\tanb$ vs.\ $\sina$ plane, including contours of constant $\cos(\beta\pm\alpha)$ and $\sbpa$. 
In all four plots, the red points have $\mha\le\mhh/2$ while 
the blue points have $\mhl\le\mhh/2$. 
Note that there are no red points for Type~II; moreover, there are no $\cbpa\sim 1$ points in Type~II that pass all constraints.
}
\label{Hfig1}
\end{figure}

Putting everything together, including also the experimental constraints, we end up with the situation shown 
in \Fig{Hfig1}. The top row shows allowed points  in the $m_{12}$ vs.\ $\tan\beta$ plane (analogous to \Fig{pertfigH}); 
the bottom row displays these same  allowed  points in the $\tanb$ vs.\ $\sina$ plane. 
As explained at the beginning of this section, in Type~I either $h$ or $A$ can be light (but not both) while 
in Type~II only $h$ can be light but not $A$.  To distinguish these two cases, points with $\mha<\mhh/2$ 
are shown in red and points with $\mhl<\mhh/2$ in blue.  
Considering first the top row of plots we see that, 
in agreement with \Fig{pertfigH}, there is a small allowed region with $\mhl<\mhh/2$ at $m_{12}\simeq 60$--80~GeV 
and $\tan\beta\lsim 2$. This region occurs for both Type~I and Type~II, although it is more constrained 
in Type~I (because of combined SUP+STU constraints). 
In Type~I there is moreover a diagonal strip of allowed points with $\mha<\mhh/2$ at $\tan\beta\simeq 2-12$, as expected 
from \Fig{pertfigH}. The points below this strip are mostly  $\cbpa\sim1$ points for which cancellations occur, 
cf.\ the lower-left plot of \Fig{Hfig1}; 
they can have  $\mha<\mhh/2$ or $\mhl<\mhh/2$.  Note that no such points survive in Type~II. 
Last, but not least, it is worth noting that, in contrast to the $h125$ case, in the $H125$ case there are no allowed points with 
``wrong sign'' Yukawa couplings, \ie\ points for which the couplings of the $H$ to vector bosons and to 
bottom quarks  have opposite signs.

\begin{figure}[t]
\begin{center}
\includegraphics[width=0.5\textwidth]{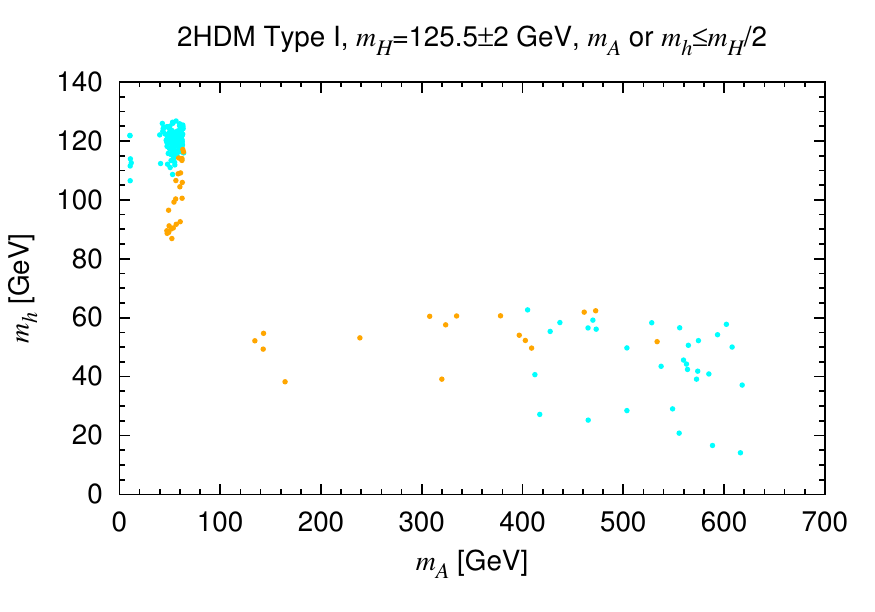}
\end{center}\vspace*{-7mm}
\caption{Allowed $H125$ points for the \typei\ model in the $\mhl$ vs.\ $\mha$ plane. 
The cyan points have $\sina>0$, while the orange points have $\sina\sim -1$, cf.\ the bottom-left plot in \Fig{Hfig1}.
\label{H125mhma}}
\end{figure}

In \Fig{H125mhma}, we take a closer look at the allowed points in the $\mhl$ vs.\ $\mha$ plane for Type~I. We see that indeed no points survive in the region where both $\mhl$ and $\mha$ are below $\mhh/2$.  As $\mhl$ increases,  some low $\mha$ points appear, but these correspond to either $\mha\lsim 12\gev$ for which there are no published limits at large $\mhl$ on $e^+e^-\to Z^*\to \hl\ha$ or to $\mha\gsim 40-50\gev$ and $\mhl\gsim 90\gev$ \ie\ sufficiently close to LEP threshold as to escape limits on the $\hl\ha$ final state by virtue of suppressed cross section. 
In the gap from about $15\gev$ to about $40\gev$, LEP limits are strong enough to eliminate all points.
It is also worth noting that the cyan points with $\sina>0$ and the orange points with $\sina\sim -1$
occupy rather distinct parts of the $\mhl$ vs.\ $\mha$  plane. In particular, if a light scalar with $\mhl<60\gev$ 
plus a pseudoscalar with $\mha<400\gev$ were discovered, this would fix $\sina\sim -1$ in Type~I.

\begin{figure}[t!]
\begin{center}
\includegraphics[width=0.5\textwidth]{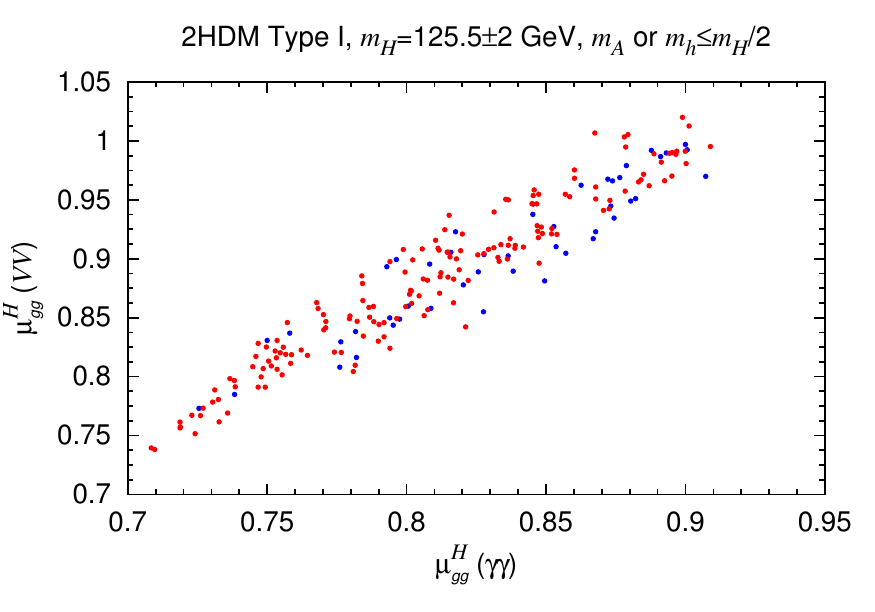}\hskip-.1in
\includegraphics[width=0.5\textwidth]{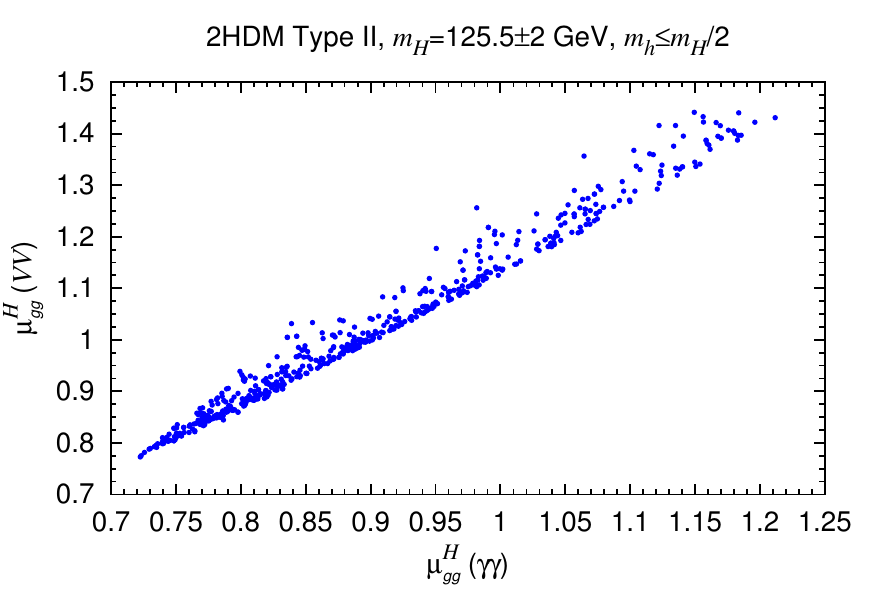} 
\end{center}\vspace*{-10mm}
\caption{Signal strengths $\rggHzz$ vs. $\rggHgamgam$ for the \typei\ and \typeii\ models. 
Points with $\mha\le\mhh/2$ are shown in red and points  with $\mhl\le\mhh/2$ in blue.}
\vskip-.1in
\label{Hmuratios}
\end{figure}

\begin{figure}[t!]
\begin{center}
\includegraphics[width=0.5\textwidth]{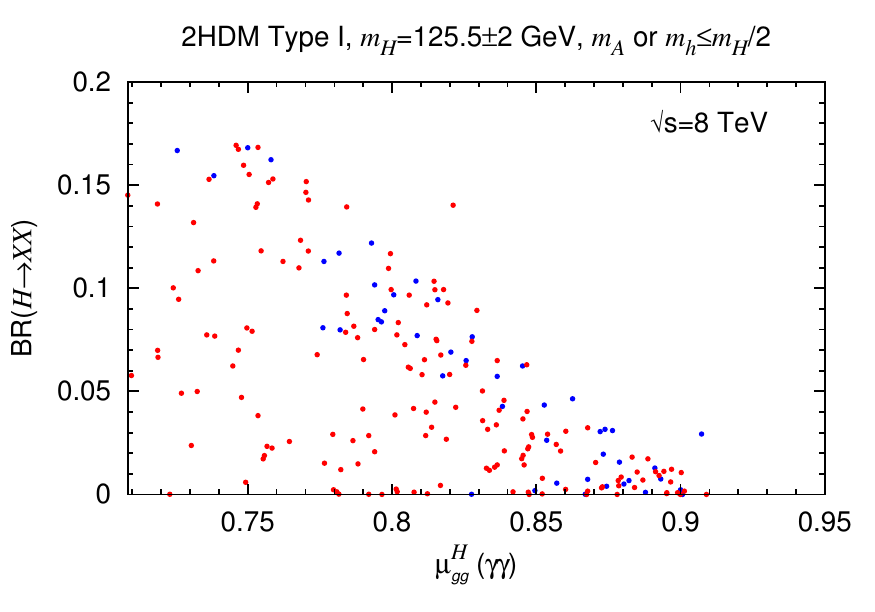}\hskip-.1in
\includegraphics[width=0.5\textwidth]{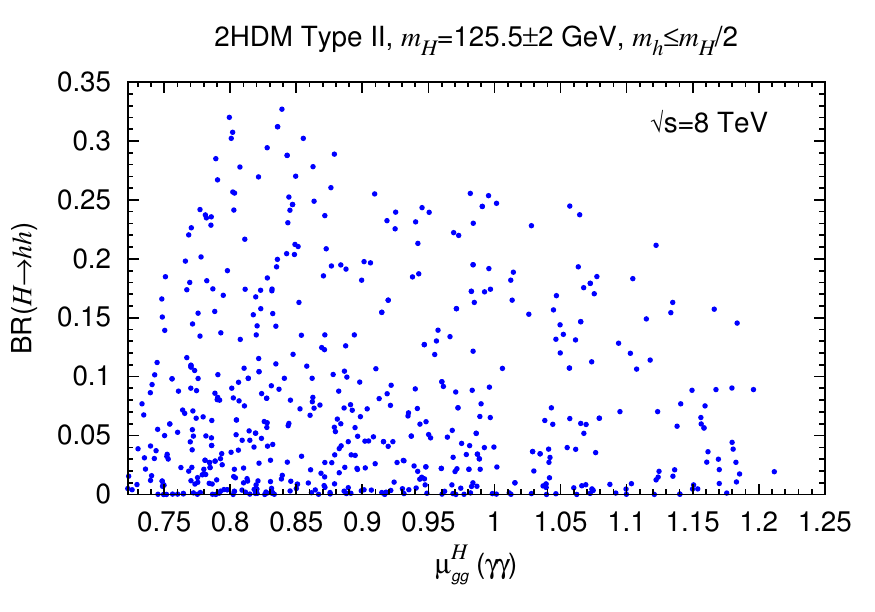} 
\end{center}\vspace*{-10mm}
\caption{Branching ratios of $H\to XX$ ($X=h,A$) decays vs.\ $\rggHgamgam$ 
for the \typei\ and \typeii\ models.  Points with $\mha\le\mhh/2$ are shown in red and points  with $\mhl\le\mhh/2$ in blue.
}
\vskip-.1in
\label{HmuBRHaa}
\end{figure}

Let us now explore the phenomenological consequences of the $H125$ scenario for the LHC. 
To this end, we first show in \Fig{Hmuratios} the relation between the signal strengths for 
the high-resolution channels $gg\to H\to VV$ ($VV=WW^{(*)},ZZ^{(*)}$) 
denoted as $\rggHzz$ and $gg\to H\to \gam\gam$ denoted as $\rggHgamgam$. 
As in the $h125$ case, quite substantial deviations from the SM values of unity are possible. 
With the increased precision expected at Run~2, the Higgs measurements at the LHC 
should be sensitive to such deviations. Moreover, also as in the $h125$ case, the exact SM 
case $\rggHgamgam=\rggHzz=1$ cannot be obtained in the $H125$ scenarios with 
light $h$ or $A$.  Though not shown here, this tension with SM-like signal strengths is also 
apparent in the $\rvbfHgamgam$ vs.\ $\rggHgamgam$ plane.
Should the signal strength measurements for either of these  pairs converge to values that lie within 10\% of their SM values the $H125$ scenarios with $m_h$ or $m_A$ below $m_H/2$ will be excluded.\footnote{Comparing 
with Fig.~7 of \cite{Dumont:2014kna} we see that this tension with SM-like signal strengths 
is much less in the general $H125$ case with heavier $h,A$.} 
For completeness we show in \Fig{HmuBRHaa} also $\br(H\to XX)$, $X=h~\textrm{or}~A$, versus 
$\rggHgamgam$. Despite the existing Run~1 constraints, the branching ratios can be sizeable and it may thus be interesting to look for these decays. 

\begin{figure}[t!]
\begin{center}
\includegraphics[width=0.5\textwidth]{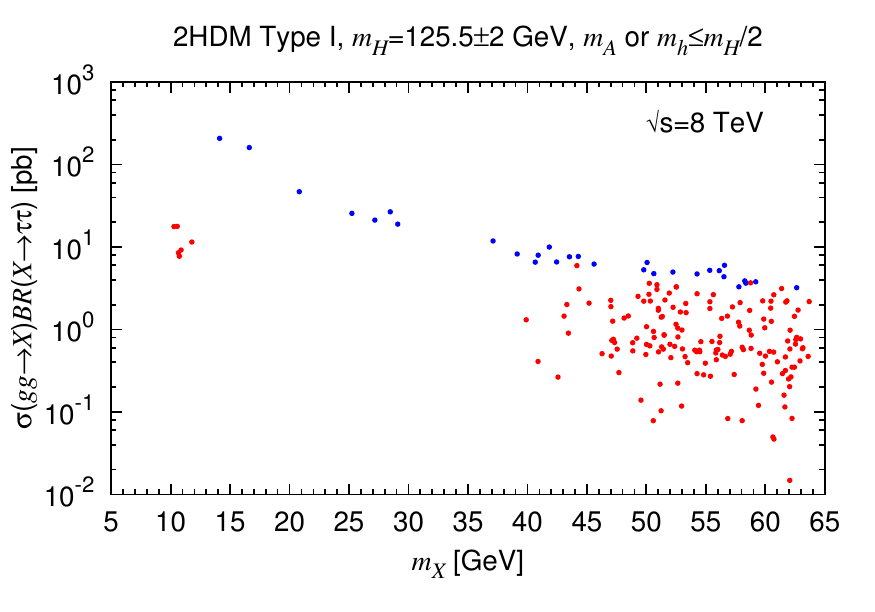}\hskip-.1in
\includegraphics[width=0.5\textwidth]{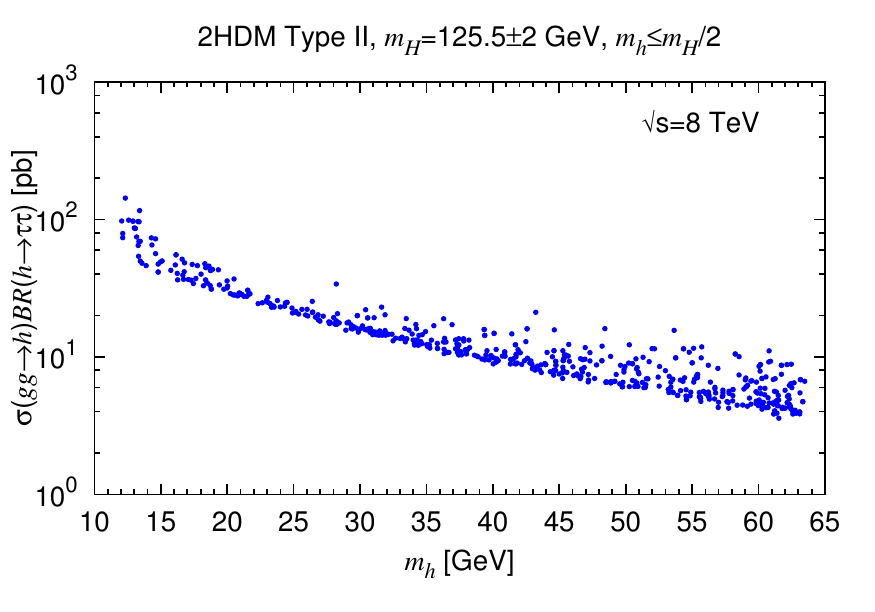}\\
\includegraphics[width=0.5\textwidth]{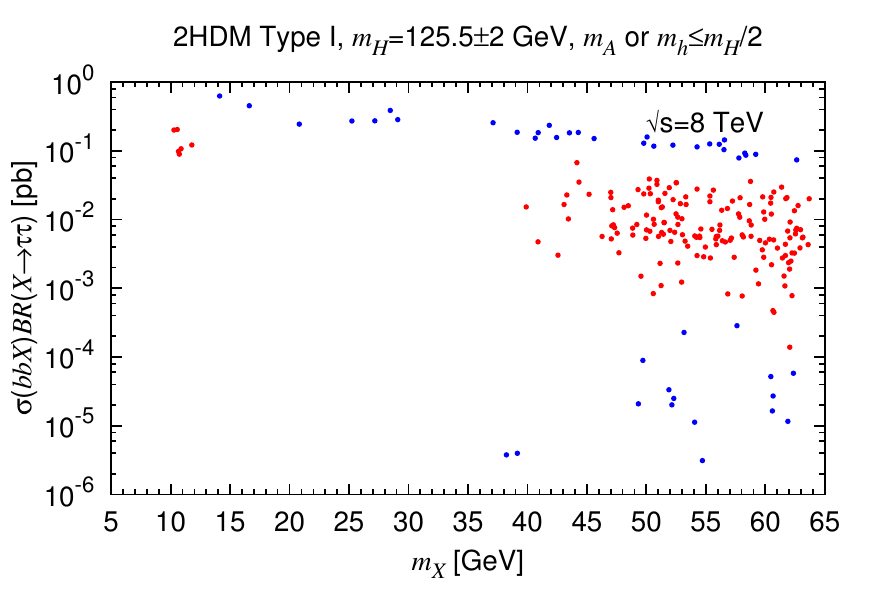}\hskip-.1in
\includegraphics[width=0.5\textwidth]{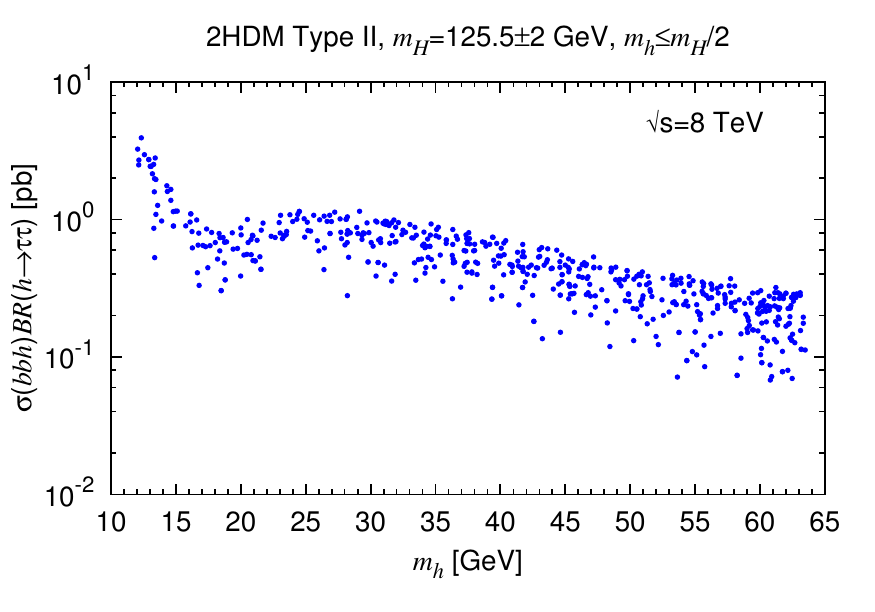}
\end{center}\vspace*{-10mm}
\caption{For the $H125$ case, we give 8 TeV cross sections for light $X=h,A$ production from $gg$ fusion (upper row) and $b\anti b$ associated production (lower row) in the $\tau\tau$ final state. The blue points are for $X=h$, the red points for $X=A$.}
\label{Hxsecsh}
\end{figure}

The most important issue is whether or not the existing 8 TeV, $L=20\fbi$ data set could be sensitive to this scenario by looking for the light $\hl$ or $\ha$ in the $\tau\tau$ or $\mu\mu$ final states. 
The relevant plots are given in \Fig{Hxsecsh}. Since $\tan\beta$ cannot be large in the \typeii\ model (see \Fig{Hfig1}) 
and there is no $\tanb$ enhancement of the $b\anti b$ coupling in the \typei\ model, it is mostly $gg$ fusion that's relevant. $\sigma(gg\to X)\times\br(X\to\tau\tau)$ exceeds the 
required 10~pb (or $0.1$~pb for decays into $\mu\mu$) in particular for the light $h$ case, $X=h$.
Light pseudoscalars (possible only in \typei) have smaller cross sections and will be harder to detect. 
Concretely, only for $gg$ fusion with $\ha\to\tau\tau$ and $\mha\lsim 12\gev$ does one obtain a cross section 
as large as $10\pb$ in the $\tau\tau$ channel, though for $\mha>40\gev$ cross sections are still between 
$1\pb$ and $10\pb$.

A final comment concerns the issue of vacuum stability in these scenarios. 
According to  \cite{Barroso:2013awa}, the 2HDM minimum is the global minimum only if 
$D\equiv m_{12}^2(m_{11}^2-k^2m_{22}^2)(\tan\beta-k)>0$, where $k=(\lambda_1/\lambda_2)^{1/4}$.  However, given that $D<0$ may still correspond to a metastable vacuum, we have chosen not to require $D>0$; one would need to 
compute the corresponding vacuum lifetime, which is beyond the scope of the present study. We note that were we to require $D>0$ this would eliminate only a small percentage of the  $h125$ scenario points, but would exclude about 
20\% of the points in the $H125$ scenario.  
We leave further investigation of the implications of   
vacuum (meta)stability to future work. 

\section{Conclusions}

We have considered 2HDM scenarios of Type~I and Type~II in which the $\ha$ or $\hl$ has mass 
below one-half that of the observed $125\gev$ SM-like Higgs state, when the latter is identified with 
either the lighter CP-even $\hl$ or heavier CP-even $\hh$. 
It turns out that this is a region which LEP limits do not constrain at all in the $h125$ case or only partially 
constrain in the $H125$ case. The conditions and associated parameter choices for obtaining viable scenarios that have a small enough 
decay branching ratios of the $\sim125\gev$ Higgs boson into a pair of lighter Higgs states were discussed 
in detail. 

Regarding LHC phenomenology, we found that in the scenarios under consideration the signal strengths of the  
$\sim125\gev$ Higgs boson cannot all be SM-like. Should the signal strength measurements in the high-resolution 
$\gam\gam$ and $VV$ channels converge to their SM values to within 10\% or better, then these scenarios will be excluded. 
Moreover, in the $h125$ case, surprisingly large $gg$ fusion and $b\anti b$ associated production cross sections are possible for a light pseudoscalar in the 10--60~GeV mass range; naive estimates suggest that these should be readily testable in the $\tau\tau$ and $\mu\mu$ channels using the existing 8 TeV data from Run~1 of the LHC. 

Overall, one finds ample motivation from these 2HDM scenarios for the ATLAS and CMS collaborations to explore their sensitivity to Higgs particles with masses below about $60\gev$ in the $\tau\tau$ and $\mu\mu$ final states.  If sufficient sensitivity is reached and nothing is observed, then many of the 2HDM scenarios explored in this paper will be eliminated.  
On the other hand, if such a light Higgs is detected then models such as the MSSM will be eliminated and a strong preference in favour of, \eg, a general 2HDM or the NMSSM will arise.

\section{Acknowledgements}

This work was supported in part by US DOE grant DE-SC-000999 and by the French ANR project {\sc DMAstroLHC}. 
J.B.\ is supported by the ``Investissements d'avenir, Labex ENIGMASS''.
Y.J.\ is supported by  LHC-TI fellowship  US NSF grant PHY-0969510;  
he also thanks the LPSC Grenoble for hospitality and the Labex ENIGMASS for financial support 
for a research stay during which this work was finished.

\bibliographystyle{JHEP}
\bibliography{2hdmlowmA_v2}

\end{document}